\documentclass[fleqn,usenatbib, onecolumn]{mnras}

\usepackage{newtxtext,newtxmath}
\usepackage[T1]{fontenc}

\DeclareRobustCommand{\VAN}[3]{#2}
\let\VANthebibliography\thebibliography
\def\thebibliography{\DeclareRobustCommand{\VAN}[3]{##3}\VANthebibliography}

\usepackage{graphicx}	
\usepackage{amsmath}	
\usepackage{mathtools}
\usepackage{orcidlink}
\usepackage[caption=false,lofdepth,lotdepth]{subfig} 

\newcommand{\be}{\begin{equation}}
\newcommand{\ee}{\end{equation}}
\newcommand{\bary}{\begin{eqnarray}}
\newcommand{\eary}{\end{eqnarray}}

\graphicspath{{Figures/}}

\title[Kilonova - GRBs]{Inverse Compton scattering occurring in a reverse-shock scenario involving a kilonova: A channel of TeV gamma-ray photons}

\author[N. Fraija et al.]{
Nissim Fraija\orcidlink{0000-0002-0173-6453},$^{1}$\thanks{E-mail: nifraija@astro.unam.mx}, J. A. Montes$^{1}$\orcidlink{0000-0001-6026-7338}, Sara Fraija-Castellanos\orcidlink{0009-0008-8311-6290},$^{1}$ and Mar\'ia Magdalena Gonz\'alez$^{1}$\orcidlink{0000-0002-5209-5641}\\
$^{1}$Instituto de Astronom\' ia, Universidad Nacional Aut\'onoma de M\'exico, Circuito Exterior, C.U., A. Postal 70-264, 04510 Cd. de M\'exico, M\'exico.\\
}
\date{Accepted XXX. Received YYY; in original form ZZZ}

\pubyear{2015}

\begin{document}
\label{firstpage}
\pagerange{\pageref{firstpage}--\pageref{lastpage}}
\maketitle

\begin{abstract}

Gamma-ray bursts (GRBs) are among the most luminous transients in the Universe and constitute prime targets for multimessenger studies, particularly in connection with gravitational-wave events. The detection of very-high-energy (TeV) photons from GRBs would provide valuable constraints on the physical conditions in the outflow, including the bulk Lorentz factor, circumburst density, radiation processes, and microphysical parameters. The possible detection of TeV emission temporally associated with an optical–infrared kilonova (KN), as suggested for GRB 160821B, presents a challenge to standard synchrotron self-Compton scenarios. In this work, we explore an alternative mechanism in which TeV photons are produced during the afterglow phase via external inverse Compton (EIC) scattering. In this scenario, electrons accelerated in the reverse shock upscatter seed photons originating from the KN.  We derive the corresponding EIC light curves and spectra for a reverse shock evolving in the thin-shell regime within a constant-density medium, and apply the model to GRB 160821B. We further constrain the parameter space for TeV detectability by incorporating the high KN luminosity observed in AT2017gfo, as well as flux upper limits reported by H.E.S.S. and HAWC.  We find that TeV emission is more likely under conditions of very low magnetic energy fraction, $\epsilon_{\rm B_r} \lesssim 10^{-6}$, combined with a bright KN and relatively low redshift. This mechanism predicts TeV photons on timescales of hours to a few days after the burst.
\end{abstract}




\section{Introduction}
Gamma-ray bursts (GRBs) emit around $10^{49} - 10^{55}$~erg in pulses of gamma rays, making them the most energetic transient phenomena in the Universe. They could be produced by merging binary compact objects (BCO); a black hole (BH) - a neutron star (NS) or NS-NS \citep{1992ApJ...392L...9D, 1992Natur.357..472U, 1994MNRAS.270..480T, 2011MNRAS.413.2031M} or the collapse of massive stars \citep{1993ApJ...405..273W,1998ApJ...494L..45P, Woosley2006ARA&A}.   These exotic sources are classified as short or long, depending on the duration of the prompt gamma-ray episode ($T_{90}$).\footnote{$T_{90}$ is defined as the period during which the cumulative count of gathered data above background increases from 5\% to 95\%.}   The merger of BCOs that generate a kilonovae (KNe) \citep[e.g., see ][]{2017ApJ...848L..16S, 2017Sci...358.1556C, 2017ApJ...848L..17C}, is associated to short GRBs \citep[sGRBs; $T_{90}\lesssim 2\,{\rm s}$;][]{1998ApJ...507L..59L, 2005ApJ...634.1202R, 2010MNRAS.406.2650M, 2013ApJ...774...25K, 2017LRR....20....3M}, and the collapse of dying colossal stars that leads to a supernova (SN) \citep[e.g., see][]{1999Natur.401..453B, 2006ARA&A..44..507W}  are linked to long GRBs \citep[lGRBs; $T_{90}\gtrsim 2\,{\rm s}$;][]{1993ApJ...413L.101K}.  Regardless of the progenitor classification, observations have definitively demonstrated that highly relativistic and collimated outflows are responsible for afterglow phenomena \citep{2002ApJ...571..779P, 2015PhR...561....1K} often interpreted within the fireball model \citep[e.g. see][]{1978MNRAS.183..359C}.   This model determines the reverse \citep{2000ApJ...545..807K, 2003ApJ...597..455K, 2016ApJ...818..190F} and a forward \citep{1995ApJ...455L.143S, 1998ApJ...497L..17S, 1999ApJ...513..669K} shock when a relativistic jet launched by the progenitor sweeps the surrounding medium and transfers a substantial fraction of its energy to it. Shocked-accelerated electrons can be cooled down by synchrotron, synchrotron self-Compton (SSC), and/or external inverse Compton (EIC). For instance, in the compact-binary-merger scenario, \cite{2022Natur.612..236M} reported a significant detection of high-energy gamma-ray emission (in the $0.1-1~\rm GeV$ range) performed by Fermi-LAT in GRB 211211A, starting approximately $10^3\,{\rm s}$ after the trigger time. This emission lasted around $20,000\,{\rm s}$ and was explained by an EIC process in which the accelerated electrons in the forward-shock region upscatter KN photons.\\ 

KNe arise from the radioactive decay of heavy r-process nuclei synthesized in the neutron-rich ejecta expelled during the merger \citep{2017LRR....20....3M}. They typically produce rapidly evolving thermal emission that, for a few days, can be detected in the ultraviolet, optical, and near-infrared bands. The brightness and color evolution of the light curve of a KN (initially more “blue” and then transitioning to “red”) are considered a signature of the formation and content of these heavy elements. Observationally, KNe serve as a robust electromagnetic counterpart to short GRBs, confirming the origin of some sGRBs in neutron star mergers \citep{2010MNRAS.406.2650M, 2015MNRAS.446.1115M}. Their discovery and study have provided fundamental insights into the astrophysical sites of r-process nucleosynthesis and the origin of many of the heaviest elements in the Universe \citep{2019LRR....23....1M}. Disentangling the properties of KNe is an important point especially given the association of sGRBs with the Gravitational Waves (GWs). The detection of GRB 170817A,  AT 2017gfo and GW170817 has paved the way on the nature of sGRBs as the coalescence of NS mergers. The evident KN signature in GW170817 provided a chance to estimate their detectability in sGRBs and the variability in their features.   Since sGRBs are usually discovered via detection of the $\gamma$-ray prompt emission from the relativistic jet, they are typically observed where the afterglow is the brightest and thus most probably obscure the KN. This emission with its more isotropic component is easier to see at angles far away from the sGRB jet \citep{2012ApJ...746...48M}. Despite this, it has been possible to determine only five KNe. The bursts  associated with the claimed KNe  are GRB 050709 \citep{2016NatCo...712898J}, GRB 060614 \citep{2015NatCo...6.7323Y}, GRB 130603B \citep{2013Natur.500..547T, 2013ApJ...774L..23B}, GRB 160821B  \citep{2017ApJ...843L..34K, 2019MNRAS.489.2104T} and GRB 211211A \citep{2022Natur.612..228T}. \cite{2018ApJ...860...62G} analyzed a sample of 23 short nearby GRBs ($z \leq 0.5$) to compare the optical and near-IR light curves with  AT 2017gfo.    They considered short bursts, following the historical classification, the ones with $T_{90} \leq 2$ s and also the class of sGRBs with extended emission \citep{2006ApJ...643..266N, 2010ApJ...722L.215D, 2016ApJ...825L..20D, 2017ApJ...848...88D, 2017A&A...600A..98D}. This comparison enables us to characterize their diversity in terms of their brightness distribution.   \cite{2018ApJ...860...62G}  found that for four sGRBs: 050509B, 051210, 061201, and 080905A, a KN of the same brightness of AT 2017gfo might have been observed. For these bursts, the deep 3 $\sigma$ upper limits, two times or more dim than the detections of AT 2017gfo at comparable rest-frame times, seem to exclude the presence of a KN like AT 2017gfo. The authors also found that the afterglows in GRBs 150424A, 140903A and 150101B  were too bright for an AT 2017gfo-like KN to be detected. Finally, they reported that the host galaxies of sGRBs 061006, 071227 and 170428A were to bright, and in six bursts there was no sufficient constraining observations regarding the presence of KN.\ Covering 14 years of operations with Swift, \cite{2020MNRAS.492.5011D} presented a systematic search for sGRBs in the local Universe. The authors found no events at distance $\lesssim 100\,{\rm Mpc}$ and four candidates located at  $\lesssim 200\,{\rm Mpc}$.  They derived, in each case, constraints of optical upper limits on the onset of a ``blue"  KN,  implying low mass ejecta ($\lesssim 10^{-3}\,M_\odot$).\\


The synchrotron process is considered the cooling mechanism up to a few GeVs \citep{1995ApJ...455L.143S, 1998ApJ...497L..17S, 1999ApJ...513..669K, 2000ApJ...545..807K, 2003ApJ...597..455K, 2016ApJ...818..190F}, whereas SSC and EIC models are invoked to explain the VHE photons \citep{2001ApJ...548..787S,2001ApJ...546L..33W, 2001ApJ...559..110Z, 2012ApJ...755...12V, 2019ApJ...883..162F, 2019ApJ...879L..26F, 2021ApJ...918...12F, 2024MNRAS.527.1884F, 2024MNRAS.527.1674F, 2003ApJ...598L..11G,2001ApJ...546L..33W,2001ApJ...556.1010W}.  As electrons are accelerated in external shocks, VHE photons are expected to be delayed relative to low-energy photons emitted in the prompt episode.   It should be noted that VHE photons are not expected during the prompt phase due to $\gamma\gamma$-opacity effects \citep[e.g., see][]{2006ApJ...650.1004B}.  The TeV emission observed on Earth is expected to come from the closest and strongest bursts due to interactions with low-energy photons from the extragalactic background light \citep[EBL;][]{1966PhRvL..16..252G}.  To far, there have been six bursts with VHE photons ($\geq 100\,{\rm GeV}$; GRB 160821B, 180720B, 190114C, 190829A, 201216C and 221009A) detected by MAGIC \citep{2019Natur.575..459A, 2021ApJ...908...90A}, H.E.S.S. \citep{2019Natur.575..464A, 2021Sci...372.1081H} and the Large High Altitude Air Shower Observatory \citep[LHAASO;][]{2023arXiv230606372L} in temporal coincidence with the afterglow. For instance, GRB 180720B and GRB 190829A were detected by H.E.S.S. Telescopes ten and four hours, respectively, after the trigger time, GRB 190114C was detected for almost 40 minutes by the MAGIC telescope beginning one minute post-trigger, and GRB 221009A was observed by LHAASO 230 s after the trigger.

In this paper, we propose a model to explain the presence of TeV photons during the afterglow phase considering the EIC scenario; shock-accelerated electrons were accelerated during the reverse shock interacting with the seed photon from the KN.  Keeping this in mind, the following is the structure of the paper: In Section \S\ref{sec2} we derive the relevant equations of our model. In Section \S\ref{sec3}, we apply the theoretical scenario to model the sub-TeV observation reported in GRB 160821B and explore the parameter space so that the TeV emission was detected in GRB 170817A, and finally, in Section \S\ref{sec5}, we provide a summary of our work and some concluding remarks.  We will adopt the convention $Q_{\rm x}=\frac{Q}{10^{\rm x}}$ in cgs units and assume for the cosmological constants a spatially flat universe $\Lambda$CDM model with  $H_0=69.6\,{\rm km\,s^{-1}\,Mpc^{-1}}$, $\Omega_{\rm M}=0.286$ and $\Omega_\Lambda=0.714$ \citep{2016A&A...594A..13P}.


\section{External Inverse Compton Scenario}\label{sec2}
\subsection{Synchrotron Reverse-shock model: Thin-shell regime}
When most of the total energy from the relativistic outflow is transferred to the circumburst medium, 
a reverse shock that propagates back into the jet
could occur \citep{1997ApJ...476..232M, 1999ApJ...517L.109S, 2000ApJ...542..819K}.  The amount of total energy delivered to the reverse-shock region is redistributed to the electrons and the comoving magnetic field ($B'_r$) via the microphysical parameters $\epsilon_{e_r}$ and $\epsilon_{B_r}$, respectively. Unprimed quantities denote measurements taken in the observer's frame, while primed quantities refer to those in the comoving frame. Consequently, relativistic electrons embedded within a magnetic field undergo acceleration and cooling by the synchrotron process. The comoving magnetic field $B'_r=\sqrt{8\pi}\,{U'}_{\rm B_r}^{\frac12}$ is calculated from the magnetic energy density ($U'_{\rm B_r}$).  We analyze the synchrotron reverse-shock radiation emitted from both on-axis and off-axis jets, which is decelerated within a constant-density medium.   Since we focus on a short-lived component temporarily separated from the main emission, we assume the reverse shock operates in the thin-shell regime, where the time needed for shock crossing is longer than the duration of the prompt emission.  The effects in the self-absorption regime are not addressed, as their contribution is generally substantial only at very low energies \citep[e.g., see][]{2014ApJ...788...70P}.

\subsubsection{On-axis Synchrotron scenario}

 The dynamics of the reverse shock is determined by the Sedov length \citep{1995ApJ...455L.143S}
\be\label{Sedov_on}
\ell_0=\left[\frac{3E_{\rm K}}{4 \pi m_{p} c^2\,n\,}\right]^{\frac{1}{3}}\,,
\ee

 and the width of the shell $\Delta_{\rm x}\approx c (1+z)^{-1}\,t_{\rm x}$ via the bulk Lorentz factor 
\be\label{Gamma_on}
\Gamma=\left[\frac{\ell_0\,}{\Delta_{\rm x}}\right]^{\frac{3}{8}}\,,
\ee

where $n$ is the density of the circumburst medium, $z$ is the redshift, $E_{\rm K}$ is the isotropic-equivalent kinetic energy, $c$ is the speed of light, $m_p$ the proton mass, and $t_{\rm x}$ is the shock crossing time defined by \citep{2007ApJ...655..391K, 2020ApJ...905..112F}

\be 
t_{\rm x} = \left[\frac{3\,E_{\rm K} (1+z)^{3}}{4\pi\, m_pc^{5} n\Gamma^{8}}  \right]^{\frac{1}{3}}.
\ee

The dynamics of the reverse shock is classified in the thin-shell regime when the critical Lorentz factor value is higher than the bulk Lorentz factor ($\Gamma <\Gamma_{\rm c}$) or the shock-crossing time is longer than the burst duration ($T_{90} < t_{\rm x}$). The critical Lorentz factor derived from Eqs. (\ref{Sedov_on}) and (\ref{Gamma_on}) can be written as \citep{2007ApJ...655..973K, 2003ApJ...595..950Z}

\be 
\Gamma_{\rm c}\equiv\left[\frac{3\,E_{\rm K}\,(1+z)^{3}}{4\pi\, m_p c^{5}\,n}  \right]^{\frac{1}{8}} T^{-\frac{3}{8}}_{90}\,.
\ee


Now, we estimate the dynamics of synchrotron emission before and after the shock-crossing time.  The evolution of the hydrodynamic quantities in the shocked region, such as the density, Lorentz factor, the total number of electrons, and the pressure, is introduced in, e.g., \cite{2000ApJ...545..807K, 2015AdAst2015E..13G}.

\paragraph{Before shock-crossing time ($t<t_{\rm x}$)}  Electrons, when exposed to acceleration during reverse shocks, display a helical motion inside enhanced magnetic fields, which inevitably results in the emission of photons through the synchrotron mechanism. Given an electron distribution ($\gamma_{\rm e}^{\rm -p} d\gamma_{\rm e}$ for $\gamma_{\rm m,r}\leq\gamma_e$ with $p$ the spectral index), the minimum and the cooling Lorentz factors before the reverse shock passes through the shell are

{\small
\begin{eqnarray}
\gamma_{\rm m, r} &\simeq& 
1.3\times 10\,\left(1+z\right)^{-3}   \epsilon_{\rm e_r} \chi_{\rm e}^{-1}\Gamma^{8}_{2} n_{-1} E^{-1}_{\rm K, 51}\,t^{3}_2 
\cr
\gamma_{\rm c, r}&\simeq& 5.4\times 10^5\,\left(1+z\right)  \left(1+Y_{\rm r_s} + Y_{\rm r_{ext}} \right)^{-1}\epsilon^{-1}_{\rm B_r,-4} n^{-1}_{-1}\, \Gamma^{-3}_{2} \,t^{-1}_2,
\eary
}

respectively,\footnote{In this section, we will use a spectral index value of $p=2.2$ and a redshift of $z=0.1$ to estimate each proportionality constant unless specified otherwise.}  where $Y_{\rm r_s}=\frac{U'_{\rm syn}}{U'_{\rm B}}$, $\chi_{\rm e}$ is the fraction of the shocked-accelerated electrons and $Y_{\rm r_{ext}}=\frac{U'_{\rm ext}}{U'_{\rm B_r}}$ are the Compton parameters, $U'_{\rm syn}$ the synchrotron photon energy density and $U'_{\rm ext}=L_{\rm KN}/(4\pi R^2c\Gamma^2)$ the external photon energy density of the KN with $L_{\rm KN}$  the KN luminosity at a particular frequency ($\nu_{\rm seed}$) and $R$ is the shocked radius. The characteristic and cooling spectral breaks of synchrotron emission ($\nu^{\rm syn}_{\rm \rm m/c,r}\propto \gamma_{\rm m/c,r}^2 B'$) and the maximum flux density are given by  

{\small
\begin{eqnarray}\label{break_thin_bef_on}
h \nu^{\rm syn}_{\rm m, r} &\simeq& 
2.1\times 10^{-5}\,{\rm eV}\,\left(1+z\right)^{-7}   \epsilon^2_{\rm e_r,-1} \epsilon^{\frac12}_{\rm B_r,-4} \chi_{\rm e}^{-2} \Gamma^{18}_{2} n_{-1}^{\frac52} E^{-2}_{\rm K, 51}\,t^{6}_2 \cr
h\nu^{\rm syn}_{\rm c, r}&\simeq& 3.8\times 10^4\,{\rm eV}\,\left(1+z\right)  \left(1+Y_{\rm r_s} + Y_{\rm r_{ext}} \right)^{-2}\epsilon^{-\frac32}_{\rm B_r,-4} n_{-1}^{-\frac32}\, \Gamma^{-4}_{2} \,t^{-2}_2,\cr
F^{\rm syn}_{\rm max,r} &\simeq& 5.9\times 10^{2}\,{\rm mJy} \,\left(1+z\right)^{-\frac12} \epsilon^{\frac12}_{\rm B_r,-4}\, \chi_{\rm e}\,n_{-1} \, d^{-2}_{\rm z,27.1}\,\Gamma^{5}_{2}\, E^{\frac12}_{\rm K, 51}\,t^{\frac32}_2\,,\,\,\,\,\,\,\,\,
\eary
}

respectively, where the term $d_{\rm z}$ corresponds to the luminosity distance for a flat $\Lambda CDM$ universe, which is calculated as $d_{\rm z}=(1+z)\frac{c}{H_0}\int^z_0 \frac{d\tilde{z}}{\sqrt{\Omega_{\rm m}(1+\tilde{z})^3 + \Omega_{\rm \Lambda}}}$ \citep[e.g., see][]{1972gcpa.book.....W}.   The maximum energy radiated by the synchrotron mechanism in the reverse-shock region is computed by comparing the acceleration and synchrotron timescales. Therefore, the maximum Lorentz factor of the electron distribution is $\gamma_{\rm max,r}=\left(3q_e/\xi\sigma_T B'_{\rm r}\right)^{\frac12}$  and the maximum synchrotron energy is

\begin{equation}
h \nu^{\rm syn}_{\rm max, r}=2.3\,{\rm GeV} \left(1+z\right)^{-1}\, n^{-\frac18}_{-1} E^{\frac{1}{8}}_{\rm K, 51} \,t^{-\frac{7}{16}}_{2}\,.
\end{equation}

The term $\xi\approx 1$ corresponds to the parameter in the Bohm limit, $\sigma_T$ the cross section in the Thompson regime and $q_e$ the elementary electron charge. Using the synchrotron spectral breaks and the maximum flux (eq. \ref{break_thin_bef_on}), the synchrotron light curves in the fast- and slow-cooling regime can be written as

{\small
\begin{eqnarray}\label{ThinAfterSyncSlow0}
F^{\rm syn}_{\nu,r}\propto
\begin{cases}
t^{\frac{13}{6}}  \, \nu^{\frac13},\hspace{1.8cm} \nu<\nu^{\rm syn}_{\rm m,r},\cr
t^{\frac{1}{2}}\nu^{-\frac{1}{2}},\hspace{1.7cm} \nu^{\rm syn}_{\rm m,r}<\nu<\nu^{\rm syn}_{\rm c,r},\,\,\,\,\,\cr
t^{\frac{6p-5}{2}}\nu^{-\frac{p}{2}},\hspace{1.3cm}\nu^{\rm syn}_{\rm c,r}<\nu\,, \cr
\end{cases}
\end{eqnarray}
}

and

{\small
\begin{eqnarray}\label{ThinAfterSyncSlow0}
F^{\rm syn}_{\nu,r}\propto
\begin{cases}
t^{-\frac{1}{2}}  \, \nu^{\frac13},\hspace{1.8cm} \nu<\nu^{\rm syn}_{\rm m,r},\cr
t^{\frac{6p-3}{2}}\nu^{-\frac{p-1}{2}},\hspace{1.1cm} \nu^{\rm syn}_{\rm m,r}<\nu<\nu^{\rm syn}_{\rm c,r},\,\,\,\,\,\cr
t^{\frac{6p-5}{2}}\nu^{-\frac{p}{2}},\hspace{1.3cm}\nu^{\rm syn}_{\rm c,r}<\nu\,, \cr
\end{cases}
\end{eqnarray}
}
respectively.

\paragraph{After shock-crossing time ($t_{\rm x} < t$)}
Once the reverse shock has traversed the shell,
the minimum and the cooling Lorentz factors are

{\small
\begin{eqnarray}
\gamma_{\rm m, r} &\simeq& 
1.1\times 10\,\left(1+z\right)^{\frac27}  \epsilon_{\rm e_r,-1} \chi_{\rm e}^{-1}\Gamma^{-\frac{16}{21}}_{2} n_{-1}^{-\frac{2}{21}} E^{\frac{2}{21}}_{\rm K, 51}\,t^{-\frac{2}{7}}_3 \cr
\gamma_{\rm c, r}&\simeq& 2.5\times 10^6\,\left(1+z\right)^{-\frac{19}{35}}  \left(1+Y_{\rm r_s} + Y_{\rm r_{ext}} \right)^{-1}\epsilon^{-1}_{\rm B_r,-4} n^{-\frac{17}{35}}_{-1}\, \Gamma^{\frac{39}{35}}_{2}E^{-\frac{18}{35}}_{\rm K, 51} \,t^{\frac{19}{35}}_3\,,
\eary
}
respectively. At energies exceeding $h\nu^{\rm syn}_{\rm c,r}$ with $h$ the constant Planck, the synchrotron flux diminishes entirely, as there are no longer any shocked electrons contributing to it. This significant decline highlights the critical interplay between electron behavior and synchrotron emissions. Therefore,  the fluid expands adiabatically and the cutoff energy ($h\nu^{\rm syn}_{\rm cut,r}$) instead of cooling energy ($h\nu^{\rm syn}_{\rm c,r}$) \citep{2000ApJ...542..819K}.   The synchrotron spectral breaks and the maximum flux density are given by

{\small
\begin{eqnarray}\label{break_thin_aft_on}
h \nu^{\rm syn}_{\rm m, r} &\simeq& 
1.8\times 10^{-6}\,{\rm eV}\,\left(1+z\right)^{\frac{19}{35}}  \epsilon^2_{\rm e_r,-1} \epsilon^{\frac12}_{\rm B_r,-4} \chi_{\rm e}^{-2}\Gamma^{-\frac{74}{35}}_{2} n_{-1}^{-\frac{1}{70}} E^{\frac{18}{35}}_{\rm K,51}\,t^{-\frac{54}{35}}_3 \cr
h\nu^{\rm syn}_{\rm cut, r}&\simeq& 2.9\times 10^{2}\,{\rm eV}\,\left(1+z\right)^{\frac{19}{35}}  \left(1+Y_{\rm r_s} + Y_{\rm r_{ext}} \right)^{-2}\epsilon^{-\frac32}_{\rm B_r,-4} n^{-\frac{283}{210}}_{-1}\, \Gamma^{-\frac{292}{105}}_{2}E^{-\frac{16}{105}}_{\rm K, 51} \,t^{-\frac{54}{35}}_3,\cr
F^{\rm syn}_{\rm max,r} &\simeq& 8.5\times 10\,{\rm mJy} \,\left(1+z\right)^{\frac{69}{35}} \epsilon^{\frac12}_{\rm B_r,-4}\,\chi_{\rm e}\,n_{-1}^{\frac{37}{210}} \, d^{-2}_{\rm z,27.1}\,\Gamma^{-\frac{167}{105}}_{2}\, E^{\frac{139}{105}}_{\rm K, 51}\,t^{-\frac{34}{35}}_3\,,\,\,\,\,\,\,\,\,
\eary
}

respectively. During this time interval, the maximum synchrotron energy evolves as 

\begin{equation}
h \nu^{\rm syn}_{\rm max, r}=1.1\,{\rm GeV} \left(1+z\right)^{-\frac43}\, n^{-\frac19}_{-1}\,\Gamma^{\frac19}_2\, E^{\frac{1}{9}}_{K,51}  \,t^{-\frac{1}{3}}_{\rm 3}\,.
\end{equation}

Using the synchrotron spectral breaks and the maximum flux (eq. \ref{break_thin_aft_on}), the synchrotron light curves in the slow-cooling regime can be written as

{\small
\begin{eqnarray}\label{ThinAfterSyncSlow0}
F^{\rm syn}_{\nu,r}\propto
\begin{cases}
t^{-\frac{16}{35}}  \, \nu^{\frac13},\hspace{1.8cm} \nu<\nu^{\rm syn}_{\rm m,r},\cr
t^{-\frac{27p+7}{35}}\nu^{-\frac{p-1}{2}},\hspace{0.9cm} \nu^{\rm syn}_{\rm m,r}<\nu<\nu^{\rm syn}_{\rm cut,r},\,\,\,\,\,\cr
0,\hspace{2.6cm}\nu^{\rm syn}_{\rm cut,r}<\nu\,. \cr
\end{cases}
\end{eqnarray}
}

It is crucial to understand that synchrotron emission will stop dramatically when the observed energy band surpasses the energy break  ($\nu^{\rm syn}_{\rm cut,r} <\nu$) for a period longer than the shock crossing time.  This occurs because there will no longer be any shocked electrons to contribute  to the observed radiation.  The pronounced steep decay in the emission is anticipated, primarily resulting from geometric effects such as curvature. These effects govern both the decay and the predictable spectral softening, as described by $F^{\rm syn}_{\rm \nu, r}\propto t^{-(\beta+2)}\nu^{-\beta}$ with $\beta=\frac{p}{2}$ \citep{2000ApJ...543...66P, 2003ApJ...597..455K, 2004ApJ...614..284D, 2015ApJ...808...33U}.  For this case, the closure relations for the synchrotron reverse-shock scenario are  $F^{\rm syn}_{\rm \nu, r}\propto t^{-\frac{4(17+27\beta)}{70}}$ for 
$\nu^{\rm syn}_{\rm m,r}<\nu<\nu^{\rm syn}_{\rm cut,r}$ and $\propto t^{-(\beta+2)}$
for $\nu^{\rm syn}_{\rm cut,r}<\nu$.

\subsubsection{Off-axis Synchrotron scenario}


In this scenario,  the angle-dependent Sedov length  can be written as \citep{2025arXiv250212757A}

\be\label{sedov_en}
\ell(\Delta\theta)=\left[\frac{3E_{\rm K}(\Delta\theta)}{4 \pi m_{p} c^2\,n\,}\right]^{\frac{1}{3}}\,,
\ee

and the bulk Lorentz factor as

\be\label{Gamma}
\Gamma(\Delta\theta)=\left[\frac{\ell(\Delta\theta)}{\Delta_{\rm x}}\right]^{\frac{3}{8}}\,,
\ee

where $E_{\rm K}(\Delta \theta)= E_{\rm K}\,\left(\frac{1+(\Gamma\Delta\theta)^2}{2\Gamma^2}\right)^{3}\approx E_{\rm K}\,\Delta\theta^6 $ and $\ell(\Delta\theta)=\ell_0(1+(\Gamma\Delta\theta)^2)\approx \ell_0 \Delta\theta^2$ with $\ell_0$ given in Eq. (\ref{Sedov_on}).  We use the width of the shell $\Delta_{\rm x}\approx c (1+z)^{-1}\,t_{\rm x}$  and the Doppler factor $\delta_D=\frac{1}{\Gamma(1-\mu\beta_{\rm \Gamma})}$, with the approximation $(1-\beta_{\rm \Gamma}\cos(\Delta\theta))^{-1}\approx \frac{2\Gamma^2}{1 + (\Gamma\Delta\theta)^2}$,  where  $\mu=\cos \Delta \theta$ with $\Delta \theta=\theta_{\rm obs} - \theta_{\rm j}$ defined by the half-opening ($\theta_{\rm j}$) and
the viewing ($\theta_{\rm obs}$) angles, and $\beta_{\rm \Gamma}=v/c$ with $v$ the velocity of the material.   When we take into account the limit of the point source ($\theta_{\rm obs} \gtrsim 2\theta_{\rm j}$), the model demonstrates impressive precision and reliability in its results \citep[e.g., see][]{2002ApJ...570L..61G}.  It is essential to recognize that we can confidently disregard the integration over the solid angle in this approximation, allowing us to simplify our analysis and focus on the most critical aspects \citep{2017ApJ...850L..24G, 2004ApJ...614L..13E}.  It should be noted that in the relativistic case ($\beta_{\rm \Gamma}\approx 1$) when the on-axis scenario is reached ($\Delta \theta\approx 0$), the term $(1-\beta_{\rm \Gamma})^{-1}\approx2\Gamma^{2}$, and therefore the standard value of the Lorentz factor $\Gamma=\left(\ell_0/\Delta_x \right)^{\frac{3}{8}}$ is recovered \citep{1995ApJ...455L.143S, 2015AdAst2015E..13G}. 

The dynamics of the reverse shock is classified in the thick- or thin- shell regime depending on the value of the critical Lorentz factor ($\Gamma_c$) which is given by

\be 
\Gamma_{\rm c}\equiv\left[\frac{3\,E_{\rm K}(\Delta\theta)}{4\pi\, m_p c^{5}} (1+z)^{-3}  \right]^{\frac{1}{8}} T^{-\frac{3}{8}}_{90}\,.
\ee

\paragraph{Before shock-crossing time ($t<t_{\rm x}$)}

Before the reverse shock crosses the shell, it is essential to  define the minimum and cooling Lorentz factors, the synchrotron spectral breaks, and the maximum flux density. It is essential to understand these factors to accurately evaluate the dynamics involved. The minimum and cooling Lorentz factors of the electron distribution are given by

{\small
\bary \label{MagneticAndLorentz_before_thin}
\gamma_{\rm m, r}&\simeq& 4.5\times 10\, \left(1+z\right)^{-3}n_{-1}\epsilon_{\rm e_r,-1}\chi_{\rm e}^{-1}\Delta\theta^{-6}_{15}\theta_{j,5}^{2}\Gamma_{2} E_{\rm j,50}^{-1}t^{3}_2\cr
\gamma_{\rm c, r}&\simeq& 5.1\times 10^4\left(1+z\right)n_{-1}^{-1} \left(1+Y_{\rm r_s} + Y_{\rm r_{ext}} \right)^{-1}\epsilon_{\rm B_r,-4}^{-1}\Gamma^{-1}_{2}\,\Delta\theta^{2}_{15}\,t^{-1}_2 \,,
\eary
}
respectively. We adopt the notation $\Delta\theta_{15}=\frac{\Delta\theta}{15^\circ}$ and $\theta_{j,5}=\frac{\theta_{j}}{5^\circ}$ for the difference angle and half-opening angle, respectively. The synchrotron spectral breaks and the maximum flux density can be written as
{\small
\bary\label{ThinBeforeSyncSpec}
h\nu^{\rm syn}_{\rm m, r}&\simeq& 4.3\times 10^{-7}\,{\rm eV}\,\left(1+z\right)^{-7} \chi_{\rm e}^{-2}n_{-1}^{\frac{5}{2}} \epsilon_{\rm e_r,-1}^2 \epsilon_{\rm B_r,-4}^{\frac{1}{2}}\Gamma^{2}_{2}\Delta\theta^{-14}_{15}\theta_{j,5}^{4}E_{\rm j,50}^{-2}t^{6}_2\cr
h\nu^{\rm syn}_{\rm c,r}&\simeq&  4.3\times 10^{-2}\,{\rm eV}\,\left(1+z\right)n_{-1}^{-\frac{3}{2}} \left(1+Y_{\rm r_s} + Y_{\rm r_{ext}} \right)^{-2}\epsilon_{\rm B_r,-4}^{-\frac{3}{2}}\Gamma^{-2}_{2}\Delta\theta^{2}_{15}t^{-2}_2\cr 
F^{\rm syn}_{\rm max,r} &\simeq& 7.1\times 10\,{\rm mJy}\,\left(1+z\right)^{-\frac{1}{2}}n_{-1}\epsilon_{\rm B_r,-4}^{\frac{1}{2}}\chi_{\rm e} d_{\rm z,27.1}^{-2}\Gamma^{-\frac52}_{2}\Delta\theta^{-3}_{15}\theta_{j,5}^{-1}E_{\rm j,50}^{\frac{1}{2}} t^{\frac{3}{2}}_2\,.
\eary
}
Using the synchrotron spectral breaks and the maximum flux (eq. \ref{ThinBeforeSyncSpec}), the synchrotron light curves in the fast- and slow-cooling regime can be written as
{\small 
\begin{eqnarray}\label{ThinBeforeSyncFast}
\label{scsyn_t}
F^{\rm syn}_{\nu,r}\propto
\begin{cases}
t^{\frac{13}{6}}  \, \nu^{\frac13},\hspace{2.0cm} \nu<\nu^{\rm syn}_{\rm c,r},\cr
t^{\frac{1}{2}}\,\nu^{-\frac{1}{2}}\,\hspace{2.0cm}  \nu^{\rm syn}_{\rm c,r}<\nu<\nu^{\rm syn}_{\rm m,r},\,\,\,\,\,\cr
t^{\frac{6p - 5}{2}}\,\nu^{-\frac{p}{2}},\hspace{1.4cm}\nu^{\rm syn}_{\rm m,r}<\nu\,, \cr
\end{cases}
\end{eqnarray}
}
and
{\small
\begin{eqnarray}\label{ThinBeforeSyncSlow}
F^{\rm syn}_{\nu,r}\propto
\begin{cases}
t^{-\frac{1}{2}}  \, \nu^{\frac13},\hspace{1.6cm} \nu<\nu^{\rm syn}_{\rm m,r},\cr
t^{\frac{6p - 3}{2}}\nu^{-\frac{p-1}{2}},\hspace{0.9cm} \nu^{\rm syn}_{\rm m,r}<\nu<\nu^{\rm syn}_{\rm c,r},\,\,\,\,\,\cr
t^{\frac{6p - 5}{2}} \,\nu^{-\frac{p}{2}},\hspace{1cm}\nu^{\rm syn}_{\rm c,r}<\nu\,, \cr
\end{cases}
\end{eqnarray}
}
respectively.\\

%
%

\paragraph{After shock-crossing time ($t_{\rm x} < t$)}  Once the reverse shock has passed through the shell, the minimum and cooling Lorentz factors are defined as

{\small
\bary \label{MagneticAndLorentz_after_thin0}
\gamma_{\rm m, r}&\simeq& 1.9\times 10\left(1+z\right)^{\frac{2}{7}}n^{-\frac{2}{21}}_{-1} \epsilon_{\rm e_ r,-1} \chi_{\rm e}^{-1} \Gamma^{-\frac{2}{21}}_{2}\Delta\theta^{\frac{4}{7}}_{15}\theta_{j,5}^{-\frac{4}{21}}E_{\rm j,50}^{\frac{2}{21}}t^{-\frac{2}{7}}_{3} \cr
\gamma_{\rm c, r}&\simeq&2.4\times 10^{5}\,\left(1+z\right)^{\frac{9}{35}}n_{-1}^{-\frac{79}{105}} \left(1+Y_{\rm r_s} + Y_{\rm r_{ext}} \right)^{-1}\epsilon_{\rm B_r,-4}^{-1}\Gamma^{-\frac{79}{105}}_{2}\Delta\theta^{\frac{18}{35}}_{15}\theta_{j,5}^{\frac{52}{105}}E_{\rm j,50}^{-\frac{26}{105}}t^{-\frac{9}{35}}_{3} \,,
\eary
}
respectively.   At energies exceeding $h\nu^{\rm syn}_{\rm c,r}$, the synchrotron flux completely decreases, as there are no longer any shocked electrons contributing to it. The fluid undergoes adiabatic expansion, and the cutoff energy can be estimated as  $\nu^{\rm syn}_{\rm cut, r}=\nu^{\rm syn}_{\rm c, r}(t_{\rm x})\,\left(\frac{t}{t_{\rm x}} \right)^{-\frac{54}{35}}$. This notable decreases highlights the crucial connection between electron behavior and synchrotron emissions, underscoring the importance of these interactions in our understanding of their dynamics.   The synchrotron spectral breaks and maximum flux are given by
    
{\small
\bary\label{ThinAfterSyncSpec0}
h\nu^{\rm syn}_{\rm m,r}&\simeq& 2.8\times 10^{-8}\,{\rm eV}\,\left(1+z\right)^{-\frac{9}{35}}n^{\frac{53}{210}}_{-1} \epsilon_{\rm e_ r,-1}^2 \epsilon_{\rm B_r,-4}^{\frac{1}{2}}\chi_{\rm e}^{-2}\Gamma^{-\frac{26}{105}}_{2}\Delta\theta^{-\frac{18}{35}}_{15}\theta_{j,5}^{-\frac{52}{105}}E_{\rm j,50}^{\frac{26}{105}}t^{-\frac{26}{35}}_{3}\cr
h\nu^{\rm syn}_{\rm cut,r}&\simeq&  4.8\times 10^{-3}\,{\rm eV}\,\left(1+z\right)^{\frac{19}{35}}n_{-1}^{-\frac{283}{210}} \left(1+Y_{\rm r_s} + Y_{\rm r_{ext}} \right)^{-2}\epsilon_{\rm B_r,-4}^{-\frac{3}{2}}\Gamma^{-\frac{194}{105}}_{2}\Delta\theta^{\frac{38}{35}}_{15}\theta_{j,5}^{\frac{32}{105}}E_{\rm j,50}^{-\frac{16}{105}}t^{-\frac{54}{35}}_{3}\cr
F^{\rm syn}_{\rm max,r} &\simeq& 1.2\times 10^{2}\,{\rm mJy}\, \left(1+z\right)^{\frac{13}{35}}n_{-1}^{\frac{149}{210}}\epsilon_{\rm B_r,-4}^{\frac{1}{2}}\chi_{\rm e}d_{\rm z,27.1}^{-2}\Gamma^{-\frac{293}{105}}_{2}\Delta\theta^{-\frac{44}{35}}_{15}\theta_{j,5}^{-\frac{166}{105}}E_{\rm j, 50}^{\frac{83}{105}} t^{\frac{22}{35}}_{3}\,.
\eary
}
%
%
%

Using the synchrotron spectral breaks and the maximum flux (eq. \ref{ThinAfterSyncSpec0}), the synchrotron light curves in the slow-cooling regime can be written as
	%
%
%
{\small
\begin{eqnarray}\label{ThinAfterSyncSlow0}
F^{\rm syn}_{\nu,r}\propto
\begin{cases}
t^{\frac{92}{105}}  \, \nu^{\frac13},\hspace{1.8cm} \nu<\nu^{\rm syn}_{\rm m,r},\cr
t^{\frac{35-13p}{35}}\nu^{-\frac{p-1}{2}},\hspace{0.9cm} \nu^{\rm syn}_{\rm m,r}<\nu<\nu^{\rm syn}_{\rm cut,r},\,\,\,\,\,\cr
0,\hspace{2.6cm}\nu^{\rm syn}_{\rm cut,r}<\nu\,. \cr
\end{cases}
\end{eqnarray}
}
%

As the frequency band crosses the break (i.e., when \(\nu^{\rm syn}_{\rm cut,r} < \nu\)), synchrotron emission ceases (Eq. \ref{ThinAfterSyncSlow0}). The steep decay is softening due to geometric effects such as curvature, and therefore, the flux evolves as $F^{\rm syn}_{\rm \nu, r}= A_{\rm hle} t^{-(\beta+2)}\nu^{-\beta}$ with $A_{\rm hle}$ the normalized constant  \citep{2000ApJ...543...66P, 2003ApJ...597..455K, 2004ApJ...614..284D, 2015ApJ...808...33U}. For this scenario, the closure relations for the synchrotron reverse-shock scenario are  $F^{\rm syn}_{\rm \nu, r}\propto t^{-\frac{4(17+27\beta)}{70}}$ for 
$\nu^{\rm syn}_{\rm m,r}<\nu<\nu^{\rm syn}_{\rm cut,r}$ and $\propto t^{-(\beta+2)}$
for $\nu^{\rm syn}_{\rm cut,r}<\nu$.

\subsection{External Inverse Compton}

An isotropic radiation field interacts with electrons accelerated in the reverse-shock region, leading to inverse-Compton scattering to higher energies (see Figure \ref{fig:kn_jet_interac}). During the deceleration phase, an outflow that is slightly off-axis will enter our field of view (FOV) after a couple of hours, when the reverse shock is in the thin-shell regime, as expected in the KN scenario.  It is important to note that even though the relativistic outflow was initially off-axis when the isotropic KN photons interacted with the electrons a couple of hours after the trigger time, the reverse shock falls under the thin-shell case. The scattered photons produced during this interaction will appear in our field of view.   The Compton parameter allows for the inclusion of multiple contributions, which is the key to yielding light curves that incorporate both SSC and EIC.  The maximum flux resulting from the KN seed photons becomes $F^{\rm EIC}_{\rm max,r}\sim g(p)\sigma_T n\,R\, F_{\rm ext}$ where $F_{\rm ext}\sim \frac{L_{\rm seed}}{d_z^2 \nu_{\rm seed}}$ with $g(p)=\frac{p-1}{p-2}$ \citep{2008MNRAS.384.1483F, 2021ApJ...920...55Z}.  For later target photons, such as KN photons, the EIC light curve has a flatter profile compared to the SSC light curve, which dominates at later times \citep{2011ApJ...732...77M}.

\subsubsection{Before the shock crossing time ($t < t_{\rm x}$)}\label{bfc}

For this time interval, the spectral breaks and the maximum flux of the EIC emission are
\bary\label{ssc_before}
h\nu^{\rm EIC}_{\rm m, r}&\simeq& 3.8\times 10^{-3}\,{\rm eV}\, \left(1+z \right)^{-13}\,\epsilon^4_{\rm e_r,-1} \epsilon^{\frac12}_{\rm B_{r},-4}\,\chi_{\rm e}^{-4}\,n^{\frac{9}{2}}_{-1} \Gamma^{34}_{2}\, E^{-4}_{\rm K, 51}\,t^{12}_{2}\cr
h\nu^{\rm EIC}_{\rm c, r}&\simeq& 1.1\times 10^4\,{\rm TeV}\,\left(1+z \right)^{3}\,\left(1+Y_{\rm r_s} + Y_{\rm r_{ext}} \right)^{-4}  \epsilon^{-\frac72}_{\rm B_{r},-4}\,n^{-\frac{7}{2}}_{-1}\, \Gamma^{-10}_{2} \, t^{-4}_{2}\,,\cr
F^{\rm EIC}_{\rm max,r} &\simeq& 2.2\times 10^{-8}\,{\rm mJy}\,  \left(1+z \right)^{-1}\,n_{-1}\, \Gamma^{2}_{2}\, d^{-2}_{\rm z,27.1}\,t_{2}\,L_{\rm seed}\, \nu^{-1}_{\rm seed} \,.
\eary

The break energy above the Klein-Nishina regime is \citep{2011ApJ...732...77M}
{
\bary
h \nu^{\rm Klein}_{\rm c,r}&\simeq& 2.5\times 10\,{\rm TeV}\, \left(1+Y_{\rm r_s} + Y_{\rm r_{ext}} \right)^{-1}\,\epsilon^{-1}_{\rm B_{r},-4}\,n^{-1}_{-1}\,\Gamma^{-2}_{2}\,t^{-1}_{2}\,.
\eary
}

The EIC light curves before the shock-crossing time for the fast- and slow-cooling regimes are therefore given by

{
\begin{eqnarray}
\label{fast_before_thin}
F^{\rm EIC}_{\rm \nu,r} \propto
\begin{cases} 
 t^{\frac{7}{3}}\, \nu_{\rm }^{\frac13},\hspace{1.5cm} \nu<\nu^{\rm EIC}_{\rm c,r}, \cr
 t^{-1}\, \nu_{\rm }^{-\frac12},\hspace{1.4cm} \nu^{\rm EIC}_{\rm c,r}<\nu<\nu^{\rm EIC}_{\rm m,r}, \cr
 t^{6p-7}\,\nu_{\rm }^{-\frac{p}{2}},\,\,\,\,\, \hspace{0.45cm} \nu^{\rm EIC}_{\rm m,r}<\nu\,, \cr
\end{cases}
\end{eqnarray}
}
and
{
\begin{eqnarray}
\label{slow_before_thin}
F^{\rm EIC}_{\rm \nu,r} \propto
\begin{cases} 
t^{-3}\, \nu_{\rm }^{\frac13},\hspace{1.6cm} \nu<\nu^{\rm EIC}_{\rm m,r}, \cr
t^{6p-5}\, \nu_{\rm }^{-\frac{p-1}{2}},\hspace{0.7cm} \nu^{\rm EIC}_{\rm m,r}<\nu<\nu^{\rm EIC}_{\rm c,r}, \cr
 t^{6p -7}\,\nu_{\rm }^{-\frac{p}{2}},\,\,\,\,\,\hspace{0.6cm}\nu^{\rm EIC}_{\rm c,r}<\nu\,,
\end{cases}
\end{eqnarray}
}
respectively.
\subsubsection{After the shock crossing time ($t > t_{\rm x} $)}

Following the same previous process,  the spectral breaks and the maximum flux of the EIC emission are given by
{
\bary\label{ssc_before}
h \nu^{\rm EIC}_{\rm m,r}&\simeq& 2.3\times 10^{-4}\,{\rm eV} \left(1+z\right)^{\frac{39}{35}}\,\epsilon^4_{\rm e_r,-1} \epsilon^{\frac12}_{\rm B_{r},-4}\chi_{\rm e}^{-4} \,n^{-\frac{43}{210}}_{-1}\, \Gamma^{-\frac{382}{105}}_{2}\, E^{\frac{74}{105}}_{\rm K,51}\,t^{-\frac{74}{35}}_{3}\,\cr
h \nu^{\rm EIC}_{\rm cut,r}&\simeq& 2.1\times 10^3\,{\rm TeV} \left(1+z\right)^{\frac{39}{35}}\,\left(1+Y_{\rm r_s} + Y_{\rm r_{ext}} \right)^{-4}  \epsilon^{-\frac72}_{\rm B_{r},-4}\,n^{-\frac{201}{70}}_{-1} \, \Gamma^{-\frac{174}{35}}_{2}\, E^{-\frac{22}{35}}_{\rm K,51}\, t^{-\frac{74}{35}}_{3}\,\cr
F^{\rm EIC}_{\rm max,r} &\simeq&  5.6\times 10^{-8}\,{\rm mJy}\,  \left(1+z \right)^{-\frac13}\,n^{\frac79}_{-1}\, \Gamma^{\frac29}_{2}\, d^{-2}_{\rm z,27.1}\, E^{\frac{2}{9}}_{\rm K,51}\,t^{\frac13}_{2}\,L_{\rm seed}\, \nu^{-1}_{\rm seed} \,.\cr
&&\hspace{5.5cm}\eary
}

The break energy above the Klein-Nishina regime is given by \citep{2010ApJ...712.1232W}

{
\bary
h\nu^{\rm KN}_{\rm c,r}&\simeq& 5.7\times 10\,{\rm TeV}  \, \left(1+Y_{\rm r_s} + Y_{\rm r_{ext}} \right)^{-1}\,\epsilon^{-1}_{\rm B_{r},-4}\,n^{-\frac{13}{21}}_{-1} \,\Gamma^{\frac{22}{21}}_{2}\,E^{-\frac{8}{21}}_{\rm K, 51}\,t^{-\frac{1}{7}}_{3}.\,\,\,
\eary
}
In analogy to the description of the EIC light curves in section \ref{bfc} for the thin-shell regime, the EIC light curve derivation after the shock-crossing time for the slow-cooling regimes yields
%
%
%
{
\begin{eqnarray}
\label{slow_after_thin}
F^{\rm EIC}_{\rm \nu,r} \propto
\begin{cases} 
t^{\frac{109}{105}}\, \nu_{\rm }^{\frac13},\hspace{1.5cm} \nu<\nu^{\rm EIC}_{\rm m,r}, \cr
t^{\frac{146-111p}{105}}\, \nu_{\rm }^{-\frac{p-1}{2}},\hspace{0.6cm} \nu^{\rm EIC}_{\rm m,r}<\nu <\nu^{\rm EIC}_{\rm cut}, \cr
0,\,\,\,\,\,\hspace{2.1cm}\nu^{\rm EIC}_{\rm cut}<\nu\,. \cr
\end{cases}
\end{eqnarray}
}


After the flux reaches its peak ($t>t_{\rm x}$), and once the break frequency $\nu^{\rm EIC}_{\rm cut}$ has moved into the TeV band (when $\nu^{\rm ssc}_{\rm cut,r} < \nu_{\rm TeV}$),\footnote{The TeV band varies based on the energetic band being observed by TeV gamma-ray instruments, such as HAWC, HESS, LHAASO or MAGIC Telescopes.} the TeV gamma-ray emission from the reverse-shock region ceases. However, because of the angular-time delay (where emission occurs at large angles relative to the jet axis), there is no immediate disappearance of the emission. During this time, the TeV gamma-ray flux varies according to the relationship $F_\nu = A_{\rm eic} t^{-(\beta+2)}\nu^{-\beta}$ where $A_{\rm eic}$ is the normalized constant determined from TeV observations.

\subsection{Analysis of the EIC Light curves}

Figure \ref{fig:fluences} shows the EIC light curves (upper) considering four equivalent kinetic energies ($10^{50}\leq E_{\rm K} \leq 10^{53}\,{\rm erg}$) and the bolometric KN light curve (lower) from the GW170817 event.  The EIC light curves are generated by the interaction of relativistic electrons accelerated during the reverse-shock region and KN seed photons described by the light curve taken from \citet{2017arXiv171005841S}. The reverse shock evolves in the thin-shell regime, and the jet is decelerated in the constant-density medium. The values used for the microphysical parameters, the circumburst density and efficiency for a constant-density medium fall within the range suggested to generate TeV emission during the afterglow episode in a timescale of hours.\footnote{The parameter values are $z=0.009$, $\epsilon_{\rm B_r}=10^{-7.84}$, $\epsilon_{\rm e_r}=10^{-0.3}$, $\Gamma=10^{1.31}$, $n=10^{-1.03}\,{\rm cm^{-3}}$ and $p=2.05$ \citep[e.g., see][]{2014ARA&A..52...43B, Fong_2015, 2019ApJ...883..134T}.}  The bolometric KN light curve was built with the values $M_{\rm ej}=0.034\,{\rm M_\odot}$, $v_{\rm ej}=0.2\,{\rm c}$, $k=0.1\,{\rm cm^2\,g^{-1}}$ and $\beta=-1.0$. The effect of EBL absorption introduced by \cite{2012MNRAS.422.3189G} was considered.   This figure shows that the EIC emission increases dramatically up to the peak flux at the hour timescale, and then monotonically decreases. We note that as the equivalent kinetic energy increases, the EIC flux increases, the peak flux is shifted to larger timescales, and the flux is delayed. It is worth noting that the bump profile of the KN light curve is reproduced in the EIC light curves.\\

Figure \ref{fig:kilonovas} shows the EIC light curves (upper) for different bolometric luminosities of the seed photons and the profile of the bolometric KN luminosities scaled from the GW170817 event.
We scaled the KN light curve taken from \citet{2017arXiv171005841S} to
$10^{-2}$, $10^{-1}$, $1$, $10$ and $10^{2}\,L_{\rm KN}$. The parameter values used are similar to Figure \ref{fig:fluences} for $E_{\rm K}=10^{52}\,{\rm erg}$. As expected, the EIC fluxes increase as the bolometric luminosity increases, and as the bolometric emission decreases the EIC flux is delayed. It should be noted that the peak flux has a maximum value of $\sim$ 10 hours without any shifting,  but could be observed at longer times for higher KN luminosities depending on the instrument sensitivity.\\ 

Figure \ref{fig:redshifts} displays the EIC light curves for a generic source located at different distances ($0.009\leq z\leq 0.5$). The parameter values used are similar to Figure \ref{fig:fluences} for $E_{\rm K}=10^{52}\,{\rm erg}$. We note that as redshift decreases, the EIC flux increases, and the peak fluxes are slightly moved to earlier times. Additionally, the EIC flux seems to be delayed for more distant sources.
 
Figure \ref{fig:regimehists} presents histograms illustrating the favorable directions for the parameters $E_{\rm K}$, $\epsilon_{\rm B_r}$, $\epsilon_{\rm e_r}$, $\Gamma$, $n$ and $p$ evolving under the cooling conditions $\nu^{\rm EIC}_{\rm m,r}<\nu<\nu^{\rm EIC}_{\rm c,r}$ (gray) and $\nu^{\rm EIC}_{\rm m,r}<\nu^{\rm EIC}_{\rm c,r}<\nu$ (yellow) that returned fluxes above $10^{-16}\, {\rm mJy}$ at 20 TeV.\footnote{The LHAASO Collaboration published a higher flux ($\sim 6\times 10^{-11}\,{\rm erg\,cm^{-2}\, s^{-1}}$) at $17.8^{+7.4}_{-5.1}\,{\rm TeV}$ for GRB 221009A \citep{2023SciA....9J2778C}.}   We observe that the parameter defining a clear variation in flux evolution is $\epsilon_{\rm B_r}$.  Low values of the microphysical parameter $\epsilon_{\rm B,r}\lesssim 10^{-6}$, indicate that the EIC flux evolves in the slow-cooling regime.

\paragraph{Low-value of magnetic parameter for EIC flux.}
The atypical value of the magnetic microphysical parameter ($\varepsilon_B\sim  10^{-7}$) has been found in other scenarios \citep{2019ApJ...883..134T, 2009MNRAS.400L..75K, 2010MNRAS.409..226K, 2014ApJ...785...29S, 2014MNRAS.442.3147B}.  \cite{2019ApJ...883..134T} systematically examined a sample of 59 bursts reported in 2FLGC that were characterized with photon energies greater than $\ge100$~ MeV.  Based on this catalog, the authors considered and analyzed the evolution of spectral and temporal indices through the closure relations (CRs) of the synchrotron forward-shock scenario. They found that a significant fraction of GRBs in the current analysis satisfies the criteria of the synchrotron scenario, although not all exhibit this behavior. Those bursts that satisfied these CRs remained in the cooling condition $\nu^{\rm syn}_{\rm m}<\nu_{\rm LAT}<\nu^{\rm syn}_{\rm c}$, when the synchrotron model evolved in a homogeneous medium with a low value of the magnetic microphysical parameter $\varepsilon_B<10^{-7}$. \cite{2009MNRAS.400L..75K, 2010MNRAS.409..226K} analyzed the LAT-detected bursts GRB 080916C, 090510, 090902B evoking the standard synchrotron afterglow model. These bursts were adequately described with an inferred microphysical value of $\epsilon_{\rm B}\sim 10^{-7}$.  \citet{2014MNRAS.442.3147B} considered radio afterglow observations in a sample of 38 GRBs and calculated the magnetic microphysical parameter using the synchrotron afterglow model. The results indicated that 22 bursts had a value less than $\epsilon_{\rm B}\lesssim 10^{-6}$, and 7 of them with a value of $\epsilon_{\rm B_r}\lesssim 10^{-8}$. \cite{2014ApJ...785...29S} presented a systematic analysis of magnetic fields during the afterglow phase. The sample consisted of 60 and 35 GRBs, detected in the X-ray and optical bands, respectively. In particular, they found that GRB 071025 and 071112C were successfully described that required a synchrotron model with a microphysical value of $\epsilon_{\rm B}\sim 10^{-7}$ and GRB 080607 with a value of $\epsilon_{\rm B}\sim 10^{-8}$.

In afterglow scenarios, it is typically assumed that the magnetic microphysical parameter remains constant throughout the expansion of the fireball. However, the specific mechanisms that occur in relativistic shocks, such as the transfer of energy from particles and the generation of magnetic fields, are not yet fully understood. Additionally, the magnetic parameter may not be constant; some researchers suggest that this  parameter might require rapid temporal evolution \citep[$\epsilon_{\rm B}\propto t^{\rm -b}$;][]{2013arXiv1305.3689L,2013MNRAS.428..845L, 2022ApJ...931L..19S}. For instance, \cite{2013MNRAS.428..845L} examined how variations in the magnetic parameter affect the resulting spectrum for synchrotron radiation and the dominant IC process, identifying two different cases.  In the first possible situation, the turbulence gradually decays when the parameter ${\rm b}$ falls within the ranges 
${\rm -1<b<0}$ for synchrotron radiation and $b>-4/(p+1)$ for the dominant IC mechanism.  In the second case, the turbulence decays rapidly, occurring when ${\rm b<-1}$ for the synchrotron model and within the range ${\rm -3<b<-4/(p+1)}$ for the dominant IC mechanism. \cite{2013arXiv1305.3689L} required the synchrotron scenario with the evolution of the magnetic parameter in the range of $0.4\lesssim b \lesssim 0.5$ to interpret the multiwavelength afterglow observations of GRBs 090902B, 090323, 090328, and 110731A. This evolution was associated with power-law decaying microturbulence and magnetic field amplification, which may occur in relativistic collisionless shocks. \cite{2022ApJ...931L..19S} introduced the variation of the magnetic microphysical parameter to model the multiwavelength observations of GRB 190829A.  By incorporating the evolution in this parameter during the reverse shock after the shock crossing time, they were able to interpret the detailed characteristics of the observed emissions, such as the outstanding X-ray and optical peaks.  The authors found that if the magnetic energy density remained invariant, the reverse-shock emission would overestimate the late-time  observations especially at radio wavelengths.


Considering that the magnetic parameter could vary in our scenario, after the shock crossing-time the cooling Lorentz factor of the electron distribution varies as $\gamma_{\rm c,r}\propto t^{\frac{19+35b}{35}}$, the synchrotron breaks as  $\nu^{\rm syn}_{\rm m,r}\propto t^{-\frac{108+35b}{70}}$ and  $\nu^{\rm syn}_{\rm cut,r}\propto t^{-\frac{54}{35}}$, and the maximum flux as $F^{\rm syn}_{\rm max,r}\propto t^{-\frac{68+35b}{70}}$. Finally, the syncrotron light curves would evolve as $F^{\rm syn}_{\nu,r}\propto t^{-\frac{4(7+27p)+35b(p+1)}{140}}$ for $\nu^{\rm syn}_{\rm m, r} < \nu < \nu^{\rm syn}_{\rm cut, r}$ and $\propto t^{-(\beta+2)}$ for $\nu^{\rm syn}_{\rm cut, r} < \nu$.  For this case, the closure relations for the synchrotron reverse-shock scenario are  $F^{\rm syn}_{\nu,r}\propto t^{-\frac{4(17+27\beta)+35b(\beta+1)}{70}}$ for $\nu^{\rm syn}_{\rm m, r} < \nu < \nu^{\rm syn}_{\rm cut, r}$ and  $\propto t^{-(\beta+2)}\nu^{-\beta}$ for $\nu^{\rm syn}_{\rm cut, r} < \nu$.

\paragraph{The Compton parameter and the Klein-Nishina regime.}

The Klein-Nishina effect alters electron cooling rates, indirectly affecting the overall shape of the spectrum. This is particularly relevant for electrons whose up-scattered photons fall within the Klein-Nishina domain as a result of Lorentz transformations.   As the Klein-Nishina effects become significant, the behavior of the Compton parameter changes. In this scenario, the entire Compton parameter ($Y_{\rm T}=Y_{\rm r_s} + Y_{\rm r_{ext}}$) before the reverse shock crosses the shell can be calculated as \citep{2010ApJ...712.1232W, 2026MNRAS.545f1970F}

{
\begin{eqnarray}
\label{Compt_par}
Y_{\rm T}(\gamma_{\rm c,r}) [Y_{\rm T}(\gamma_{\rm c,r}) + 1]= \frac{\epsilon_{\rm e_r}}{\epsilon_{\rm B_r}}\left(\frac{\gamma_{\rm m, r}}{\gamma_{\rm c, r}}\right)^{p-2}
\begin{cases} 
1,\hspace{4.75cm} \nu_{\rm c, r}<\nu^{\rm syn}_{\rm Klein}(\gamma_{\rm c,r}), \cr
\left( \frac{\nu^{\rm syn}_{\rm Klein}(\gamma_{\rm c,r})} {\nu^{\rm syn}_{\rm c,r}} \right)^{-\frac{p-3}{2}},\hspace{2.6cm} \nu^{\rm syn}_{\rm m,r}<\nu^{\rm syn}_{\rm Klein}(\gamma_{\rm c,r}) <\nu^{\rm syn}_{\rm c,r}, \cr
 \left( \frac{\nu^{\rm syn}_{\rm m,r}}{\nu^{\rm syn}_{\rm c,r}} \right)^{-\frac{p-3}{2}} \left( \frac{\nu^{\rm syn}_{\rm Klein}(\gamma_{\rm c,r})} {\nu^{\rm syn}_{\rm m,r}} \right)^{\frac{4}{4}}  ,\,\,\,\,\,\hspace{1.3cm}\nu^{\rm syn}_{\rm Klein}(\gamma_{\rm c,r})<\nu^{\rm syn}_{\rm m,r}\,, \cr
\end{cases}
\end{eqnarray}
}

where $h\nu^{\rm syn}_{\rm Klein}(\gamma_{\rm c,r})\simeq (1+z)^{-1} m_ec^2 \Gamma \,\gamma_{\rm c,r}^{-1}$ for $\nu^{\rm syn}_{\rm m,r} < \nu^{\rm syn}_{\rm c,r}$.  To estimate the Compton parameter ($Y_{\rm T}(\gamma_{\rm c,r})$) after  the reverse shock crosses the shell, the cooling frequency ($\nu^{\rm syn}_{\rm c,r}$) must be replaced by the cutoff frequency ($\nu^{\rm syn}_{\rm cut,r}$).  To describe GRB observations above hundreds of MeV, it is necessary to determine the Lorentz factor of electrons that may unleash high-energy photons via the synchrotron process, incorporate a new spectral break, and recalculate the Compton parameter value \citep[for discussion, see ][]{2010ApJ...712.1232W}. The relative location of the synchrotron and the characteristic energies of the EIC in relation to the Klein-Nishina threshold determine the number of spectral breaks (segments) that the EIC spectrum could display, as derived by \cite{2009ApJ...703..675N, 2010ApJ...712.1232W}.

\paragraph{Synchrotron-self Compton process.}

Daytime-scale target photons, similar to KN photons, have a higher energy density than synchrotron photons; therefore, the TeV flux from the EIC scenario is expected to be dominant under the SSC model \citep{2011ApJ...732...77M}.   For completeness, we include the SSC light curves.  The SSC mechanism acts as the high-energy extension of synchrotron radiation.  Photons produced by the synchrotron reverse-shock process can be up-scattered by the same population of electrons through inverse Compton scattering via  $\nu^{\rm ssc}_{\rm m, r}\simeq \gamma^2_{\rm m, r} \nu^{\rm syn}_{\rm m, r}$  and $\nu^{\rm ssc}_{\rm c\, (cut), r}\simeq \gamma^2_{\rm c, r} \nu^{\rm syn}_{\rm c\,(cut), r}$. In this case, the maximum SSC flux can be estimated directly from the maximum synchrotron flux and optical depth ($\tau_r$) as $F^{\rm ssc}_{\rm max, r}\simeq \tau_r\,F^{\rm syn}_{\rm max, r}$ \citep[e.g., see][]{2001ApJ...548..787S, 2010ApJ...712.1232W, 2020ApJ...905..112F}.\\

For  $t<t_{\rm x}$, the characteristic and cooling breaks and the maximum density flux of the SSC mechanism evolve as $h \nu^{\rm ssc}_{\rm m, r}\propto\,t^{12}$, $h \nu^{\rm ssc}_{\rm c, r}\propto\,t^{-4}$ and $F^{\rm ssc}_{\rm max,r}\propto t^{\frac52}$, respectively. \footnote{It is worth noting that the temporal evolution of the synchrotron spectral break $\nu^{\rm syn}_{\rm m, r}\simeq \nu^{\rm ssc}_{\rm m, r}/\gamma^2_{\rm m, r} \propto t^6$ is derived in \cite{2000ApJ...545..807K}.} Taking into account the evolution of the spectral breaks and maximum flux, the SSC light curve for this time interval becomes \citep{2025MNRAS.543.1945F}
{
\begin{eqnarray}
\label{slow_after_thin}
F^{\rm ssc}_{\rm \nu,r} \propto
\begin{cases} 
t^{-\frac{3}{2}}\, \nu_{\rm }^{\frac13},\hspace{1.6cm} \nu<\nu^{\rm ssc}_{\rm m,r}, \cr
t^{-\frac{7-12p}{2}}\, \nu_{\rm }^{-\frac{p-1}{2}},\hspace{0.6cm} \nu^{\rm ssc}_{\rm m,r}<\nu <\nu^{\rm ssc}_{\rm c,r}, \cr
t^{-\frac{11-12p}{2}}\, \nu_{\rm }^{-\frac{p}{2}},\,\,\,\,\,\hspace{0.5cm}\nu^{\rm ssc}_{\rm c,r}<\nu\,. \cr
\end{cases}
\end{eqnarray}
}

In this case, the break of the Klein-Nishina regime for  $t<t_{\rm x}$ evolves as $h \nu^{\rm ssc}_{\rm Klein, c, r}\propto\,t^{-1}$.\\

For  $t_{\rm x}<t$, the evolution of characteristic and cooling spectral breaks and the maximum density flux of SSC mechanism are $h \nu^{\rm ssc}_{\rm m, r}\propto\,t^{-\frac{74}{35}}$, $h \nu^{\rm ssc}_{\rm c, r}\propto\,t^{-\frac{74}{35}}$ and $F^{\rm ssc}_{\rm max,r}\propto t^{-\frac{27}{35}}$, respectively. Taking into account the temporal variation of the spectral breaks and maximum flux, the SSC light curve for this time interval becomes

{
\begin{eqnarray}
\label{slow_after_thin}
F^{\rm ssc}_{\rm \nu,r} \propto
\begin{cases} 
t^{-\frac{1}{15}}\, \nu_{\rm }^{\frac13},\hspace{1.6cm} \nu<\nu^{\rm ssc}_{\rm m,r}, \cr
t^{-\frac{37p-10}{35}}\, \nu_{\rm }^{-\frac{p-1}{2}},\hspace{0.6cm} \nu^{\rm ssc}_{\rm m,r}<\nu <\nu^{\rm ssc}_{\rm cut, r}, \cr
0\, ,\,\,\,\,\,\hspace{2.1cm}\nu^{\rm ssc}_{\rm cut, r}<\nu\,. \cr
\end{cases}
\end{eqnarray}
}

It should be highlighted that once the break frequency $\nu^{\rm ssc}_{\rm cut}$ has surpassed the TeV band (when $\nu^{\rm ssc}_{\rm cut,r} < \nu_{\rm TeV}$), the gamma-ray emission from the reverse shock ended. However, there is no immediate disappearance of the TeV flux due to the angular-time delay effect, and the resulting gamma-ray flux would evolve as $F_\nu = A_{\rm ssc} t^{-(\beta+2)}\nu^{-\beta}$ where $A_{\rm ssc}$ is the constant to be determined from the TeV data.\\

In this case,  the spectral break in the Klein-Nishina regime for  $t<t_{\rm x}$ would evolve as $h \nu^{\rm ssc}_{\rm Klein, c, r}\propto\,t^{\frac17}$. When the Klein-Nishina effects become significant, the new segments appear in the SSC spectrum which would evolve as  $F^{\rm ssc}_{\rm \nu,r}\propto t^{\frac{11p-9}{2}}\nu^{1-p}$ for $\nu^{\rm ssc}_{\rm Klein, c, r} < \nu < \nu^{\rm ssc}_{\rm Klein, 0, r} $, $\propto t^{4p-2}\nu^{-\frac{p+1}{2}}$ for $\nu^{\rm ssc}_{\rm Klein, 0, r} < \nu < \nu^{\rm ssc}_{\rm m, r} $ and  $\propto t^{\frac{11p-5}{2}}\nu^{-\frac{3p+1}{3}}$ for $ \nu^{\rm ssc}_{\rm m, r} < \nu$, where the new spectral break $\nu^{\rm ssc}_{\rm Klein, 0, r}$ is calculated and reported in \cite{2009ApJ...703..675N}.

\subsection{Conditions for detecting TeV photons}

In the following sequence, we list the circumstances for detecting TeV photons from this scenario.\\

\subparagraph{i) A luminous KN and powerful burst situated at a significantly low redshift.}  TeV photons from a generic source experience attenuation as a function of redshift, due to the creation of electron-positron pairs in interaction with EBL photons \citep{1966PhRvL..16..252G}.  Attenuation can be quantified using the expression $\exp[-\tau_{\gamma\gamma} (z)]$ with $\tau_{\gamma\gamma}(z)$ representing the opacity. For instance, with a low-redshift of  z=0.01,  the TeV emission at 100 GeV  and 10 TeV will be attenuated by a factor of $\sim 0.99$  and $\sim 0.68$ \citep{2017A&A...603A..34F}. On the other hand, the EIC flux increases as the KN flux and the total energy released increases.    Therefore, a very small attenuation factor together with a luminous  KN and powerful burst allows us to facilitate the detection by TeV observatories.

\subparagraph{ii) An effective combination of parameters.} Due to the Klein-Nishina regime and the evolution of the maximum EIC flux,  an optimal set of parameters makes its detection more favorable.  The electron distribution at the reverse-shock region upscatters the KN photons at the Klein-Nishina regime, and therefore the EIC emission is substantially decreased above this regime. 
Then, a suitable set of values will cause the EIC flux to be below the Klein-Nishina regime.  For a timescale of hours, the EIC flux will lie in the second PL segment of the slow-cooling regimen with a characteristic hard photon index. The EIC flux evolving in the second PL segment of the slow-cooling regime increases as the circumburst density ($\propto n$) and electron equipartition parameter ($\propto \varepsilon_{\rm e}$)  increase.  Large values of these parameters enhance the detection of EIC emission.

\subparagraph{iii) TeV photons scattered during the deceleration episode.} Although photons above hundreds of MeV up to a few GeV have been observed during the prompt phase \citep{2019ApJ...878...52A},  photons above hundreds of GeV have been detected  during the deceleration episode.  Different studies  of multi-wavelength observations  have yielded results about the places where the GeV flux originates \citep[e.g. internal and external shocks; ][]{2015PhR...561....1K}. The TeV photons in  our model are scattered during the deceleration phase, where the intrinsic attenuation due to $\gamma\gamma$ interactions is much lower than unity. In our scenario, the inherent attenuation does not decrease  the TeV flux at the reverse-shock region.

\subparagraph{iv) TeV observatories with large field of view and duty cycle.} Detecting photons above hundreds of GeVs with IACTs has been a significant problem, since the time required to locate the burst exceeds the prompt episode. Despite innumerable attempts, six sightings, GRB 160821B, 180720B, 190114C, 190829A, 201216C and 221009A) detected by MAGIC \citep{2019Natur.575..459A, 2021ApJ...908...90A}, H.E.S.S. \citep{2019Natur.575..464A, 2021Sci...372.1081H} and LHAASO \citep{2023arXiv230606372L}, have been achieved, and extensive upper VHE limits have been established using these telescopes \citep[e.g. see,][]{Albert_MAGIC, Aleksi_GRB090102,  Aharonian_2009HESS, Aharonian_GRB060602B, HESS_GRB100621A, Acciari_VERITAS,  Bartoli_ARGO, GRB150323A_Abeysekara}.  TeV observatories, characterized by a large field of view and continuous operation ($\sim 100\%$ duty cycle), do not require re-pointing to the GRB location, allowing access to data preceding and after the GRB trigger, so enabling the examination of this scenario predictions.

\section{Applications: GRB 160821B and GRB 170817A}\label{sec3}

We apply our EIC scenario to describe the TeV gamma-ray detection in GRB 160821B and to explore the parameter space with the ULs derived by HAWC and HESS observatories in GRB 170817A. We explore
the parameter space because the KN luminosity in GRB 170817A was the highest recorded.

\subsection{GRB 160821B}

On 21 August 2016, the Burst Alert Telescope (BAT) onboard the Swift satellite and Fermi-GBM triggered and located GRB 160821B at 22:29:13 and 22:29:13.33 UT, respectively.  The prompt gamma-ray episode in the 15-150 keV energy band exhibited a singular and short-lived peak with a fluence and duration of $F_\gamma =(1.10 \pm 0.10)\times10^{-7}\mathrm{erg~cm^{-2}}$ and $T_{90} \approx 0.48$ s, respectively, categorizing it as sGRB ~\citep{2016GCN.19844....1P}. The X-Ray Telescope (XRT) onboard the Swift satellite commenced observations approximately 56 seconds post-trigger and monitored a diminishing X-ray afterglow over a duration of several weeks~\citep{2016GCN.19841....1S}.  Optical and radio detected the optical/NIR and radio afterglow observations were performed by the Ultraviolet/Optical Telescope (UVOT) onboard the Swift satellite, the Hubble Space telescope, ground ‐ based telescopes (NOT, WHT, GTC, Keck I), and Very Large Array (VLA) ~\citep{2021ApJ...908...90A}. GRB 160821B was located at $z \sim 0.1613$
after analyzing the spectroscopic lines ~\citep{2019MNRAS.489.2104T}. 
Fermi LAT conducted a search for high-energy $\gamma$-ray emission in the 0.1-300 GeV range, while MAGIC conducted a search for VHE particles above $>500$ GeV from GRB 160821B \citep{2021ApJ...908...90A}.   Evidence for gamma-ray photons was detected above $\sim 500$ GeV with a significance of $\sim 3\sigma$ during afterglow observations extending up to 4 hours post-burst.\\

\cite{2019MNRAS.489.2104T} displayed and analyzed  X-ray, optical/near-infrared, and radio observations of GRB 160821B. They fitted the afterglow observations with two distinct sources of emission: the synchrotron  afterglow model, resulting from the interaction of the relativistic jet with the surrounding medium, and a KN, driven by the radioactive decay of the sub-relativistic ejecta. The description of the afterglow observations indicated the presence of a reverse shock that propagated backward into the ejecta. \cite{2019ApJ...883...48L} provided a comprehensive analysis of optical and near-infrared measurements collected with the Hubble Space Telescope, and the Swift satellite of GRB 160821B and the host galaxy.  This optical and near-infrared data collection has the best-sampled KN light curve without a gravitational wave trigger. In addition, modeling the 5 GHz radio afterglow observations revealed a reverse shock with a shock-crossing time $\sim 0.1\, {\rm day}$.  Using data at various wavelengths,  \cite{2021ApJ...908...90A} examined GRB 160821B in the context of GRB afterglow theories, supposing that the detected excess events are associated with gamma-ray emission from GRB 160821B. The TeV flux could not be fully explained by using one-zone models of synchrotron self-Compton emission from the external forward shock.  \cite{FraijaCastellanos:2025T+} derived a significance map of GRB 160821B using HAWC data. Although no TeV gamma-ray excess was detected, they set a flux upper limit of 
$3.65 \times 10^{-9}\,\rm{TeV^{-1}\,cm^{-2}\,s^{-1}}$ 
considering a spectral index of $2.07$ integrated in the energy range 0.53 -- 5.99 TeV.

\subsubsection{Analysis and Results} 


Figure \ref{fig:GRB160821B} shows the EIC and SSC light curves with observations and upper limits set by the MAGIC and HAWC observatories   \citep{2021ApJ...908...90A, FraijaCastellanos:2025T+} while considering two different shock crossing times around the MAGIC observation. The left panel presents the EIC light curve for $t_{\rm x} = 1.1 \times 10^{-1}\,\mathrm{days}$, where the shock-crossing time is assumed to overlap with the MAGIC observing window. The right panel shows the corresponding light curve for $t_{\rm x} = 6.8 \times 10^{-2}\,\mathrm{days}$, assuming the crossing occurred between 0.05 and 0.07 days as proposed by \citet{2019ApJ...883...48L}. We adopt a peak bolometric kilonova luminosity of $2\times10^{42}\,\mathrm{erg\,s^{-1}}$, comparable to AT2017gfo \citep{2017arXiv171005841S}. However, the true peak luminosity may be lower, since the kilonova associated with GRB 160821B was not observed at peak and has been suggested to be dimmer and redder than AT2017gfo \citep{2019MNRAS.489.2104T}. The evolution of the light curves is consistent with the EIC flux in the slow-cooling regime $\nu^{\rm EIC}_{\rm m,r}<\nu<\nu^{\rm EIC}_{\rm c,r}$.  The cooling and characteristic spectral breaks in this case are $h\nu^{EIC}_{\rm m, r}\simeq 1.3\times 10^{-2}\,{\rm eV}$ and $h\nu^{EIC}_{\rm c, r}\simeq 2\times 10^3\,{\rm TeV}$, respectively. The fact that the shock-accelerated electrons are not in the fast-cooling regime suggests that the prompt episode starts with an evolution from the radiative to the adiabatic regime.\\

To model the MAGIC observation, we used 18,500 samples to ensure convergence of the chains. Convergence was defined by requiring each parameter to be sampled for at least 100 times its autocorrelation time. The resulting posterior distributions are shown as a corner plot in Figure~\ref{mcmc_GRB160821B_1}. For simplicity, we only show  the corner plot of our MCMC simulations for the EIC scenario applied to GRB 160821B with $t_{\rm x}=1.1\times10^{-1}\,{\rm days}$, given that best-fit values and representative lower limits across both cases are very similar. Additionally, Table~\ref{Table:ISM_Fit_2} lists the prior and posterior parameter distributions for each scenario, and for degenerate parameters we list the values used to create the light curves in Figure \ref{fig:GRB160821B}. It is important to note that the posterior distributions of $\epsilon_{\rm e_r}$, $E_{\rm k}$, and $n$ accumulate against the upper boundaries of their adopted priors and do not exhibit finite maxima or constraining within the explored parameter space. Although the MCMC chains formally converge, extending the priors to progressively larger (and even unphysical) values causes the posterior mass to shift accordingly. This behavior demonstrates that the likelihood does not provide intrinsic constraints on these parameters within the explored range. In other words, the apparent bounds arise from prior truncation rather than from the data.

For completeness, we report the corresponding 95\% lower limits in Figure \ref{mcmc_GRB160821B_1}; however, these values should be interpreted strictly as representative limits under the adopted prior ranges. They are not statistically robust, data-driven constraints, but rather prior-dependent summaries of a runaway degeneracy in which the fit quality does not deteriorate as $E_{\rm K}$, $\epsilon_{\rm e_r}$, and $n$ increase.\\

In contrast, the posterior distributions of $\Gamma$, $p$, and $\epsilon_{B_r}$ exhibit well-defined maxima and credible intervals that remain stable under prior widening tests, indicating that these parameters are meaningfully constrained by the data. The constraint on $p$ reflects the sensitivity of the spectral shape to the high-energy electron distribution, while the constraint on $\Gamma$ indicates that the normalization and timing of the EIC component depend strongly on the outflow dynamics.  
The best-fit values for $\epsilon_{B_r}$ are $3.02\times 10^{-6}$ and $2.45\times 10^{-6}$, which are consistent with the range of values obtained in some bursts \cite[e.g., see][]{2019ApJ...883..134T, 2009MNRAS.400L..75K, 2010MNRAS.409..226K, 2014ApJ...785...29S, 2014MNRAS.442.3147B}. \cite{2019ApJ...883..134T} conducted an exhaustive study of the temporal and spectral indices reported in the Second Gamma-ray Burst Catalog (2FLGC) 
and the closure relations of the synchrotron forward-shock scenario. They considered 59 bursts and showed that although the synchrotron emission model adequately explains the spectrum and temporal indices in many of them, a significant portion of bursts cannot be accurately represented by this model. Additionally, it was reported that those that fulfill the closure relations are within the slow-cooling regime, provided that the magnetic microphysical parameter is 
low enough. \\

Overall, we note that the limited dataset constrains the dynamical and spectral properties of the emitting region more robustly than the absolute energy scale of the system. The limited temporal coverage of the MAGIC observation and the lack of observations in similar energy ranges likely restrict the constraining power of the dataset, preventing the degeneracies among these parameters from being broken in the absence of additional multi-epoch or broadband observations.

\subsection{GRB 170817A}

A modest gamma-ray prompt emission of GRB 170817A detected by the INtegral and Gamma Burst Monitor (GBM) onboard of the Fermi satellite \citep{2017ApJ...848L..14G, 2017ApJ...848L..15S} was associated with a gravitational wave (GW)  event on August 17, 2017 \citep[GW170817;][]{PhysRevLett.119.161101,2041-8205-848-2-L12}.  Following the prompt episode, an extensive observational campaign was conducted in the radio, optical and X-ray bands \citep[e.g., see][and references therein]{troja2017a, 2041-8205-848-2-L12, 2017arXiv171100243K, 2018arXiv180106164D}. Several authors conducted analyzes of the non-thermal spectrum observations of GRB 170817A that were collected during the initial $\approx 900$ days following the merger. It was demonstrated that these afterglow observations were in accordance with the synchrotron afterglow model from a forward shock produced by the deceleration of an off-axis structure jet with an opening angle of approximately $5^{\circ}$.  It should be noted that  the structure jet in some phenomenological models is formed with an off-axis jet with a cocoon/shock breakout material  \citep{2017ApJ...848L...6L, 2018MNRAS.479..588G, 2019ApJ...884...71F, 2018MNRAS.479..588G, 2018ApJ...867...95H, 2019ApJ...871..200F,2021MNRAS.503.4363U}.   The best-fit value of the viewing angle reported ranges of $15^{\circ}\leq \theta_{\mathrm{obs}}\leq 25^{\circ}$ \citep{troja2017a, 2017Sci...358.1559K, 2017MNRAS.472.4953L, 2018MNRAS.478..733L, 2018ApJ...867...57R, 2017ApJ...848L..20M, 2017ApJ...848L...6L, 2018MNRAS.479..588G, 2019ApJ...884...71F, 2018MNRAS.479..588G, 2018ApJ...867...95H, 2019ApJ...871..200F}.


The KN associated with the GW170817 event and identified as AT2017gfo \citet{2017arXiv171005841S}, represents the only electromagnetic counterpart currently recognized to a GW source. This event exhibited physical parameters that align closely with the theoretical predictions of blue KN resulting from a NS merger.  The electromagnetic emission released could be described by an ejected mass of $0.04\pm 0.01\,M_\odot$, opacity of $k\leq 0.5\,{\rm cm^{2}\,g^{-1}}$ and velocity of $0.2\pm0.1\,{\rm c}$. The  PL index of the non-thermal spectrum $\beta=-1.2^{+0.3}_{-0.3}$, and the line features in the spectra indicated a nucleosynthetic source of the r-process elements.  In the optical and ultraviolet bands, it peaked less than a day after the merger and then quickly faded away.  The first observational data of AT2017gfo indicated a luminosity above $10^{42}\,{\rm erg\,s^{-1}}$ that was brighter by at least one order of magnitude than the usual KNe brightness.  Given that the post-merger product can considerably increase the peak luminosity, \cite{2018ApJ...861..114Y} and \cite{2014MNRAS.439.3916M} argued the presence of a long-lived NS. 
 
\subsubsection{Analysis and Results} 

Figure \ref{GRB_170817A} shows the upper limits of the HAWC and HESS observatories of GRB 170817A together with the light curves obtained from four sets of parameter values chosen from the excluded 6D parameter space, illustrated in Figure \ref{param_spac_GRB170817A}. The upper limit of the HAWC observatory was set in the range of 4 -- 100 TeV at $\sim8$ hours and the upper limits of the HESS telescopes were in the ranges of 0.28 -- 2.31 TeV and 0.27 -- 3.27 TeV at $\sim5$ and $\sim30$ hours, respectively. The upper left panel displays the light curve for which all upper limits are exceeded, and the upper right panel shows the light curve for which only the HAWC UL is crossed. Similarly, the lower left and right panels show the light curves for which one of the HESS upper limits are surpassed. The parameter values for each light curve are listed in Table \ref{Table:ISM_Fit}. This table shows that almost all parameters have small fluctuations around standard values; $\epsilon_{\rm e_r}\sim 0.3$, $n\sim 0.1\,{\rm cm^{-3}}$, $p\sim 2.07$, $\Gamma\sim 30$, the equivalent kinetic energy and the magnetic microphysical parameter are in the ranges $4.37\times 10^{52}\leq E_{\rm K}\leq 9.77\times 10^{54}\,{\rm erg}$, and $6.17\times 10^{-10}\leq \epsilon_{\rm B_r}\leq 4.68\times 10^{-7}$, respectively. Taking into account the typical values reported of the equivalent kinetic energy for different models $E_{\rm K}\sim 10^{52 - 53.5}\,{\rm erg}$ \citep[e.g., see][]{troja2017a,2017ApJ...848L..14G,  2022ApJ...927L..17H}, the atypical value corresponds to the magnetic microphysical parameter.  By comparison of the parameter values found after describing the multiwavelength afterglow observations of GRB 170817A (e.g., see) 
\citep[e.g., see][]{troja2017a,2022ApJ...927L..17H} with the parameter used in our EIC model (see Table \ref{Table:ISM_Fit}), we note that a lower value of the magnetic microphysical parameter would have been detected at very high energies.\\ 

Figure \ref{param_spac_GRB170817A} was generated to map how observations of H.E.S.S. and HAWC constrain the physical parameter space associated with GRB 170817A and the EIC scenario proposed in this work. The approach begins by defining a six-dimensional domain spanned by the kinetic energy, microphysical shock parameters, Lorentz factor, circumburst density, and electron spectral index. Within these bounds, we evaluated theoretical model predictions against the flux upper limits reported by HAWC and H.E.S.S. The comparison is expressed through an exclusion score $S(\theta)$, calculated as the maximum ratio between the predicted flux and the corresponding observational limit for a given parameter set $\theta$. A model is considered excluded if $S(\theta)>1$, meaning it would have produced a detectable gamma-ray signal inconsistent with the non-detections.

To effectively characterize the transition between allowed and excluded regions, we first estimate a representative upper bound ($S_{\text{max}}$) to $S(\theta)$ across the admissible space via differential-evolution. We then construct a geometric sequence of score levels that spans from the exclusion threshold $S\simeq 1$ up to this upper bound. For each score level, models close to that expected exclusion strength are identified and used as “seeds” to generate local clusters of neighboring models. These neighborhoods are densely sampled to capture how exclusion classification varies in the vicinity of approximately constant contours $S(\theta)$. All candidate models are evaluated using the same physical requirements and observational checks to ensure consistency across the entire sample.

Once stratified sampling is complete, we project the classified models onto all informative two-dimensional planes defined by pairs of physical parameters. For each panel, we compute the total number of sampled models and the number of excluded models in a grid of bins, and we display the resulting excluded fraction. The color scale therefore indicates the degree to which the parameter combinations in each bin are disfavored by the gamma-ray non-detections.  For instance, the left uppermost panel (log($E_{\rm K}$) vs log($\epsilon_{\rm B_r}$)) exhibits a pair of rejected values around $10^{51.5}\,{\rm erg}\lesssim E_{\rm K}$ and $\epsilon_{\rm B_r}\lesssim 10^{-6}$. It should be noted that the range of values for which at least one upper limit is exceeded is surpassed: the equivalent kinetic energy ($10^{51.5}\,\lesssim E_{\rm K}\lesssim 10^{54.5}\,{\rm erg}$), the circumburst density ($10^{-2}\lesssim n \lesssim 10^{-1}\,{\rm cm^{-3}}$), the magnetic equipartition parameter ($10^{-10} \lesssim \epsilon_{\rm B_r}\lesssim 10^{-6.5}$), the electron equipartition parameter ($10^{-0.7}\lesssim \epsilon_{\rm e_r}\lesssim 10^{-0.2}$), the spectral index of the electron population $2.0\lesssim p\lesssim2.2$ and the bulk Lorentz factor $5.0\lesssim \Gamma\lesssim33$.

\section{Summary}\label{sec5}

We have derived the EIC light curves which emerge from the interaction of the relativistic electrons accelerated during the reverse-shock region and KN seed photons.  We have considered that the reverse shock is evolving in the thin-shell regime and the jet is decelerated in the constant-density medium, as expected in a circunstellar environment around sGRBs. For a realistic scenario, we considered the KN seed photons described in the GW170817 event \citep{2017arXiv171005841S} and the effect of EBL absorption introduced by \cite{2012MNRAS.422.3189G}. The parameter values required, such as the microphysical parameters, circumburst density, bulk Lorentz factor, and the equivalent kinetic energy, are within the range proposed to produce TeV emission during the afterglow phase over a timescale of hours. With the required values, the EIC scenario lies in the slow-cooling regime in the cooling condition $\nu^{\rm EIC}_{\rm m, r} < \nu < \nu^{\rm EIC}_{\rm c, r}$,  reflected in a hard photon spectrum favoring TeV observations. \\

As particular cases to apply our EIC scenario, we have required the MAGIC observation with a significance of $\sim 3\sigma$ for GRB 160821B and the upper TeV limits performed by the HAWC and HESS observatories for GRB 170817A. Although MAGIC observation was performed above $0.5\, {\rm TeV}$ at $\sim 4\,{\rm h}$, the upper limits of the HAWC and HESS observatories were carried out in the range of 4 -- 100 TeV at $12\,{\rm h}$, and in the ranges of 0.28 -- 2.31 TeV and 0.27 -- 3.27 TeV at 5 and 30 hours, respectively.

The best-fit values of the constant-density medium found to explain the TeV emission in GRB 160821B suggest that the progenitor of this burst erupted in an environment with low density.  These values are consistent with the fact that sGRBs have larger offsets than lGRBs.   The spectral index values of the electron populations agree with the usual values reported for relativistic electrons cooled by synchrotron radiation after acceleration in forward shocks \citep[e.g., see][]{2022ApJ...927L..17H}. The result confirms that the gamma-ray emission at TeV was produced during the GRB afterglow.  The best-fit values of the microphysical parameter given to accelerate electrons are larger than $\epsilon_{\rm e_r}\geq 0.1$. Since the shock-accelerated electrons are not in the fast cooling regime, as shown, the prompt emission starts with a transition from the radiative to the adiabatic phase.
The best-fit values of the magnetic parameter found $\epsilon_{\rm B_r}\leq 10^{-7}$ are consistent with the values reported in other LAT-detected bursts \citep{2019ApJ...883..134T, 2009MNRAS.400L..75K, 2010MNRAS.409..226K, 2014MNRAS.442.3147B}.\\

Taking into account that the KN (AT2017gfo) associated with the GW170817 event has been the highest ever reported,  we explored the parameter space of GRB 170817A for which TeV emission would have been observed. We found that the range of values lying in the ranges of the equivalent kinetic energy ($10^{51.5}\,\lesssim E_{\rm K}\lesssim 10^{54.5}\,{\rm erg}$), the circumburst density ($10^{-2}\lesssim n \lesssim 10^{-1}\,{\rm cm^{-3}}$), the magnetic equipartition parameter ($10^{-10} \lesssim \epsilon_{\rm B_r}\lesssim 10^{-6.5}$), the electron equipartition parameter ($10^{-0.7}\lesssim \epsilon_{\rm e_r}\lesssim 10^{-0.2}$), the spectral index of the electron population $2.0\lesssim p\lesssim2.2$ and the bulk Lorentz factor $5.0\lesssim \Gamma\lesssim33$ are excluded. 
We conclude that a combination of the optimal set of parameter values would have made the detection more favorable. For instance, a very low value of the magnetic microphysical parameter.\\

As protons could be accelerated in external shocks, VHE photons might be expected from neutral pion decay products \citep[e.g.][]{1997PhRvL..78.2292W, 2013ApJ...772L...4G, 2013PhRvL.110l1101Z}.  Nevertheless, with progressively more restrictive observations, neutrinos from GRBs are more disfavored as a population, and thus hadronic models.  Following a comprehensive analysis of several years of data, the IceCube Collaboration disclosed the absence of coincidences between GRBs and VHE neutrinos \citep{2022arXiv220511410A,2012Natur.484..351A, 2016ApJ...824..115A, 2015ApJ...805L...5A}.  Based on the GRB analysis recently performed by the Gamma-ray Burst Monitor (GBM) instrument, \citet{IceCube+22GRBlimit} rejected the contribution of GRBs to the diffuse background at levels exceeding around 10\% for the early emission.

Whereas off-axis forward-shock emission could prevail in the afterglow at later phases, inverse Compton scattering in a reverse-shock scenario with seed KN photons can provide a novel detectable early signal.   This emission may provide a distinct characteristic in the initial optical afterglow, perhaps detectable as an orphan optical counterpart in the absence of a prompt gamma-ray episode. This creates new opportunities for discovering and characterizing off-axis GRB scenarios associated with KN emission.

TeV Water Cherenkov telescopes, such as HAWC, the Southern wide-field Gamma-ray Observatory \citep[SWGO;][]{2025arXiv250601786S} and the LHAASO Observatory \citep{2023Sci...380.1390L}, are renowned for their wide field of view and nearly continuous operation—achieving an approximately 100\% duty cycle. These features, which do not require re-pointing to the GRB location, allow access to data preceding and after the GRB trigger, making them valuable tools for examining predictions related to this scenario. \citet{2023ApJ...949...15D} predicts an annual detection rate of 0.6 to 3.1 sGRBs (including both on and off-axis events) within 200 Mpc. This estimate is based on detection rates from the Swift-BAT instrument (FOV $\sim 1.4$\,sr; \citealt{2004ApJ...614..284D}) and assuming a compact object merger origin. By extrapolating these rates, LHAASO-WCDA, with its a $\sim 1.7$\,sr FOV \citep{2019arXiv190502773C}, is expected to detect approximately 0.7 to 3.8 sGRBs per year.  \cite{2025A&A...697A..36L} presented an extensive analysis of KN multi-messenger prospects for future GW observatories, including the Vera Rubin Observatory, which uses revolutionary optical wide field-of-view observation techniques. This is a substantial advancement in measuring the theoretical uncertainties that influence predictions for KNe multi-messenger detections, including TeV photons.  We emphasize that due to the EBL effect \citep{1966PhRvL..16..252G}, TeV emission from a burst at redshift $z\gtrsim 0.1$ is unlikely to be detected on Earth.

\section*{Acknowledgements}

NF and MG acknowledge the financial support from UNAM-DGAPA-PAPIIT through the grants IN112525, IG101323 and IG100726, respectively.

\section*{Data Availability}

No new data was generated or analyzed to support this research.



\bibliographystyle{mnras}
\bibliography{main} 



\clearpage
\appendix

\begin{table}
    \centering \renewcommand{\arraystretch}{1.5}\addtolength{\tabcolsep}{2pt}
    \caption{
Maximum-posterior parameter sets from MCMC simulations using the EIC scenario for GRB 160821B for two shock-crossing times.\\
$\dagger$ The posteriors of $E_{\rm k}$, $n$, and $\epsilon_{\rm e_r}$ rise toward the upper prior boundary and shift under prior widening; their reported values are therefore prior-truncated. What is shown are the values used to re-create the light curves shown in Figure \ref{fig:GRB160821B}.
}
    \label{Table:ISM_Fit_2}
    \begin{tabular}{l c  c  c}
    \hline
    \hline
     GRB 160821B & Prior & $t_{\rm x_1}$ & $t_{\rm x_2}$ \\
      & & ($1.1\times 10^{-1}\,{\rm d}$) & ($6.8\times 10^{-2}\,{\rm d}$) \\ \hline \hline
     $\mathrm{log(E_K/erg)}$                & $(47, 51.5)$ & $50.33\,\,\dagger$ & $50.42\,\,\dagger$\\
     $\mathrm{log(n/cm^{-3})}$            & $$(-6,0)$$ & $-0.40\,\,\dagger$  & $-0.31\,\,\dagger$  \\     
     $\mathrm{log(\epsilon_{\rm e_r})}$      & $$(-4, -0.1)$$ & $-0.21\,\,\dagger$ & $-0.10\,\,\dagger$ \\     
     $\mathrm{log(\epsilon_{\rm B_r})}$      & $$(-8, -1.5)$$ & $-5.52_{-0.41}^{+0.45}$ & $-5.61_{-0.33}^{+0.38}$  \\         
     $\mathrm{log(\Gamma)}$           & $$(0, 3)$$& $1.24_{-0.06}^{+0.04}$  & $1.31_{-0.05}^{+0.03}$  \\         
     $\mathrm{p}$                              &  $$(2, 3)$$ & $2.09_{-0.03}^{+0.04}$ & $2.08_{-0.03}^{+0.04}$  \\ \hline\hline
    \end{tabular}
\end{table}

\begin{table}
    \centering \renewcommand{\arraystretch}{1.5}\addtolength{\tabcolsep}{2pt}
    \caption{Parameter values used to build the light curves of GRB 170817A (see Figure \ref{GRB_170817A}) using the ULs reported by HAWC and HESS observatories.}
    \label{Table:ISM_Fit}
    \begin{tabular}{l c c c c}
    \hline
    \hline
      GRB 170817A & All ULs & HAWC & HESS  & HESS  \\
       &  & (4 -- 100) TeV & (0.28 -- 2.31) TeV & (0.27 -- 3.27) TeV  \\

      \hline \hline

     $\mathrm{log(E_K/erg)}$                & 51.89 & 51.83 & 52.64 & 53.69\\
     $\mathrm{log(n/cm^{-3})}$            & -0.04  & -0.01 & -1.07 & -1.01\\     
     $\mathrm{log(\epsilon_{\rm e_r})}$      & -0.11 & -0.30 & -0.32 & -0.30\\     
     $\mathrm{log(\epsilon_{\rm B_r})}$      & -7.12 & -6.91 & -6.33 & -7.02\\         
     $\mathrm{log(\Gamma)}$           & 1.09  & 1.08  & 1.36 & 1.32\\         
     $\mathrm{p}$                              & 2.07 & 2.06 & 2.06 & 2.07\\ \hline\hline
    \end{tabular}
\end{table}


\begin{figure}
 \centering
\includegraphics[width=0.65\linewidth]{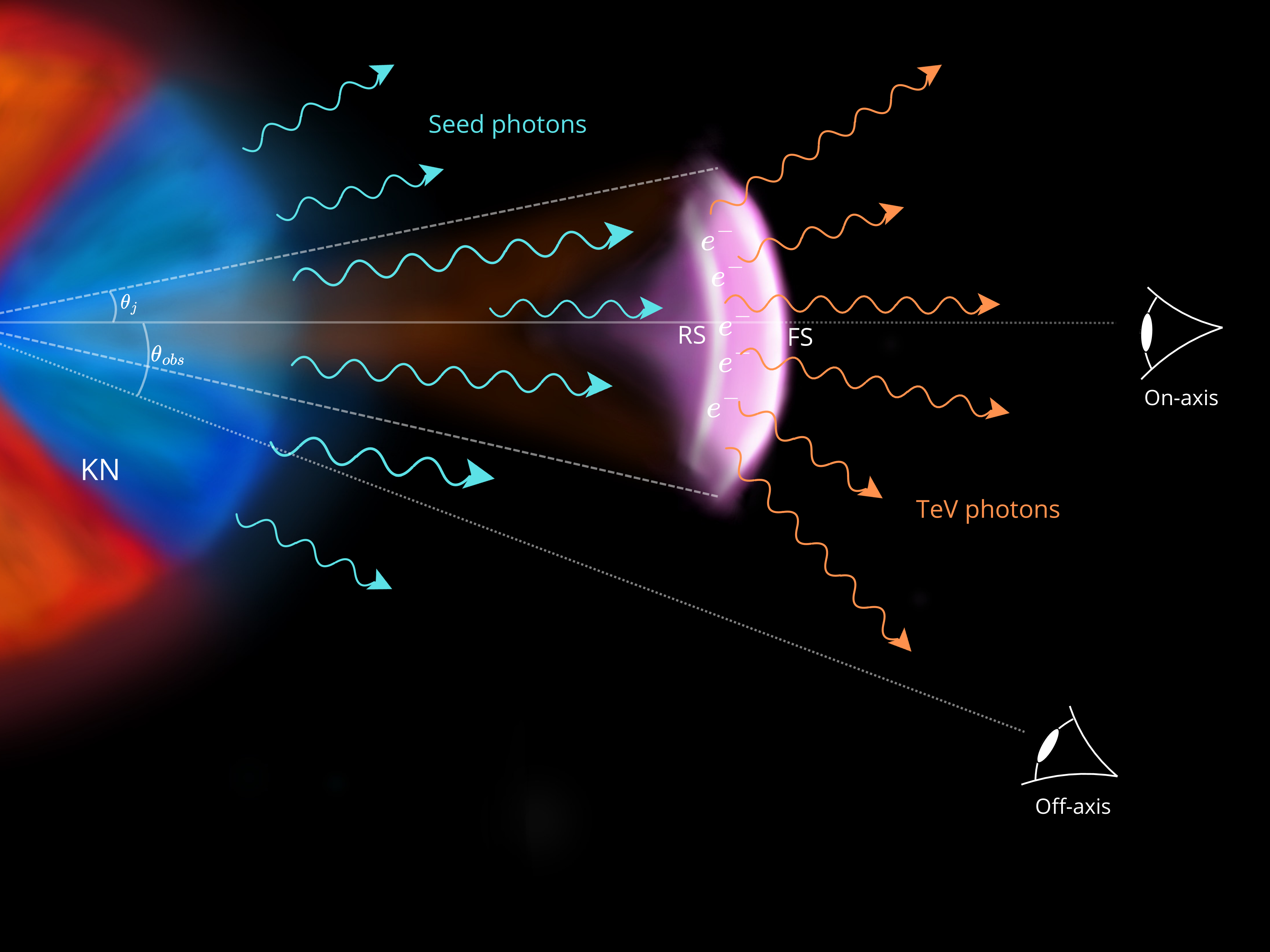}
    \caption{An isotropic radiation field originating from the KN interacts with electrons accelerated in the reverse-shock region of the jet, boosting them to higher energies via inverse Compton scattering. The on-axis and off-axis scenario is shown.}
    \label{fig:kn_jet_interac}
\end{figure}

\begin{figure}
 \centering
\includegraphics[width=0.85\linewidth]{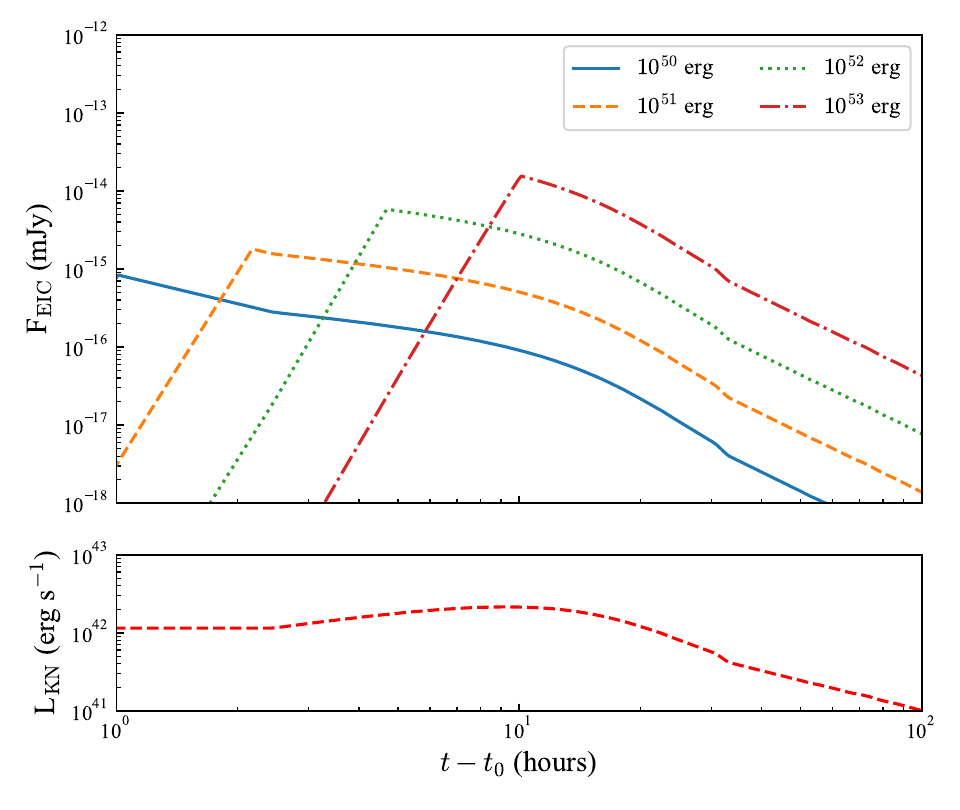} 
    \caption{Upper panel shows EIC light curves generated by relativistic electrons accelerated in the reverse-shock region and seed photons from KN considering four equivalent-kinetic energies. The reverse shock evolves in the thin-shell regime and is decelerated in the constant-density medium.  We consider the following parameters: $z=0.009$, $\epsilon_{\rm B_r}=10^{-7.84}$, $\epsilon_{\rm e_r}=10^{-0.3}$, $\Gamma=10^{1.31}$, $n=10^{-1.03}\,{\rm cm^{-3}}$ and $p=2.05$ \citep[e.g., see][]{2014ARA&A..52...43B, Fong_2015, 2019ApJ...883..134T}.   Lower panel displays the bolometric KN light curve of GW170817 taken from \citet{2017arXiv171005841S}.
    }
    \label{fig:fluences}
\end{figure}
\begin{figure}
    \centering
    \includegraphics[width=0.75\linewidth]{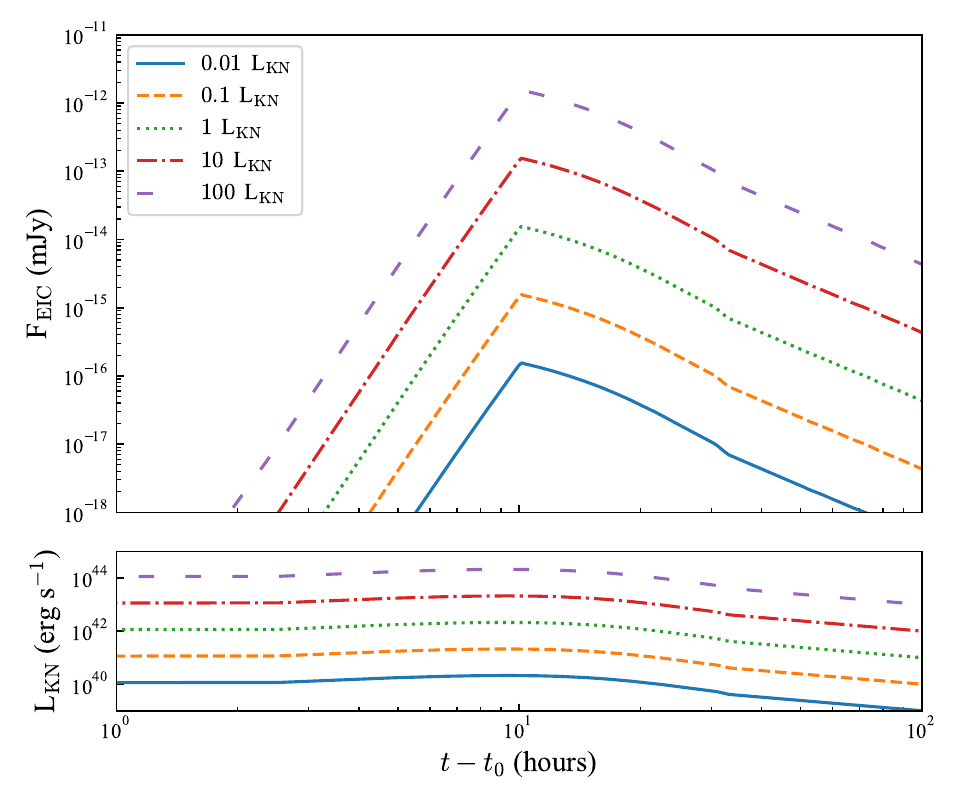}
    \caption{The same as Figure \ref{fig:fluences}, but considering the same bolometric luminosity model of seed photons at different scale factors for an equivalent-kinetic energy of ${\rm E_{\rm K}=10^{52}\,erg}$.}
    \label{fig:kilonovas}
\end{figure}
\begin{figure}
    \centering
    \includegraphics[width=0.75\linewidth]{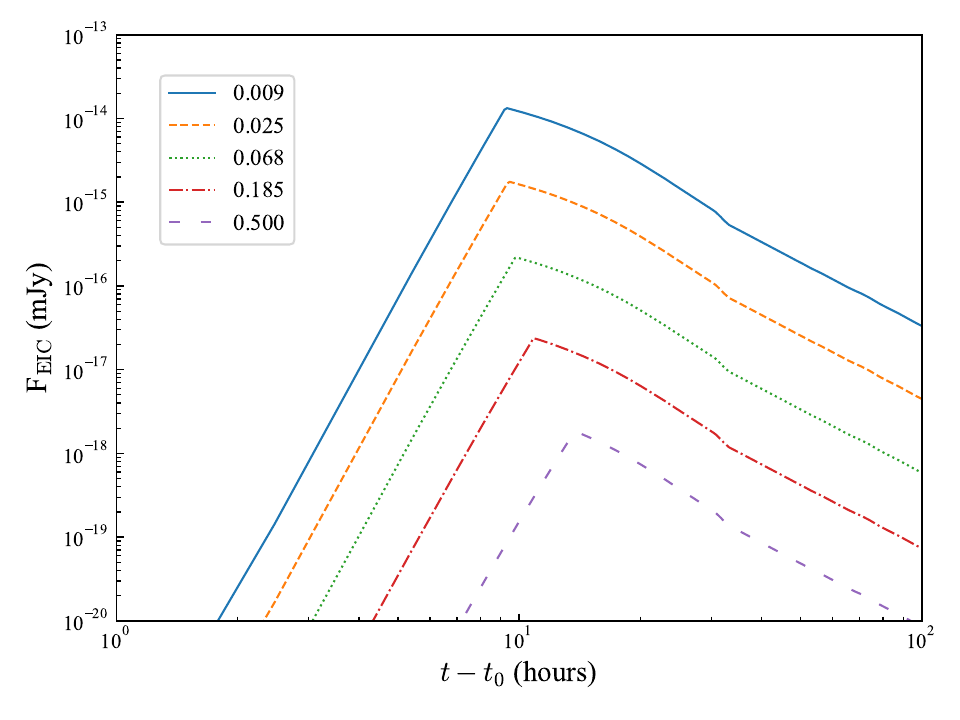}
    \caption{The same as Figure \ref{fig:fluences}, but for different redshifts and an equivalent-kinetic energy of ${\rm E_{\rm K}=10^{52}\,erg}$. }
    \label{fig:redshifts}
\end{figure}

\begin{figure}
    \centering
    \includegraphics[width=0.85\linewidth]{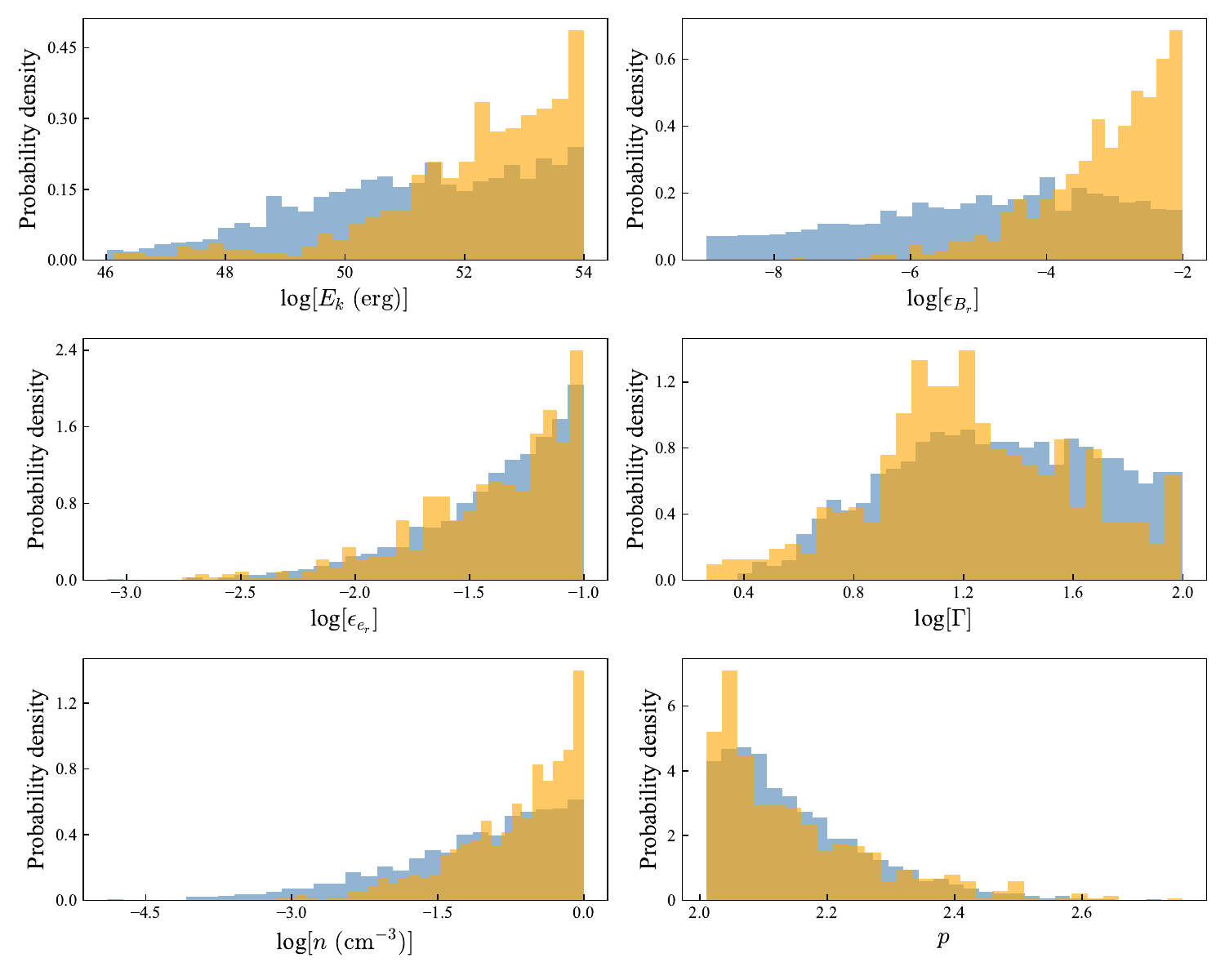}
    \caption{Histograms showcasing the favorable directions for each parameter ($E_{\rm K}$, $\epsilon_{\rm B_r}$, $\epsilon_{\rm e_r}$, $\Gamma$, $n$ and $p$)  in the large parameter combination search, corresponding to the two power laws that returned fluxes above $10^{-16}\, {\rm mJy}$ at 20 TeV.}
    \label{fig:regimehists}
\end{figure}

\begin{figure}
    \centering
    \includegraphics[width=0.47\linewidth]{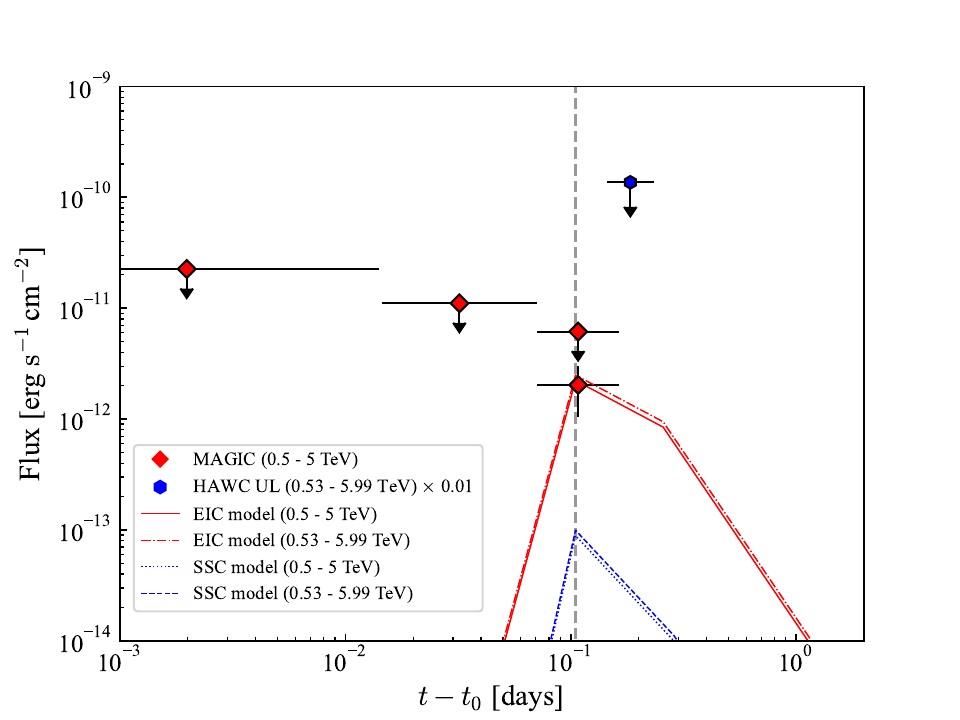}
    \includegraphics[width=0.47\linewidth]{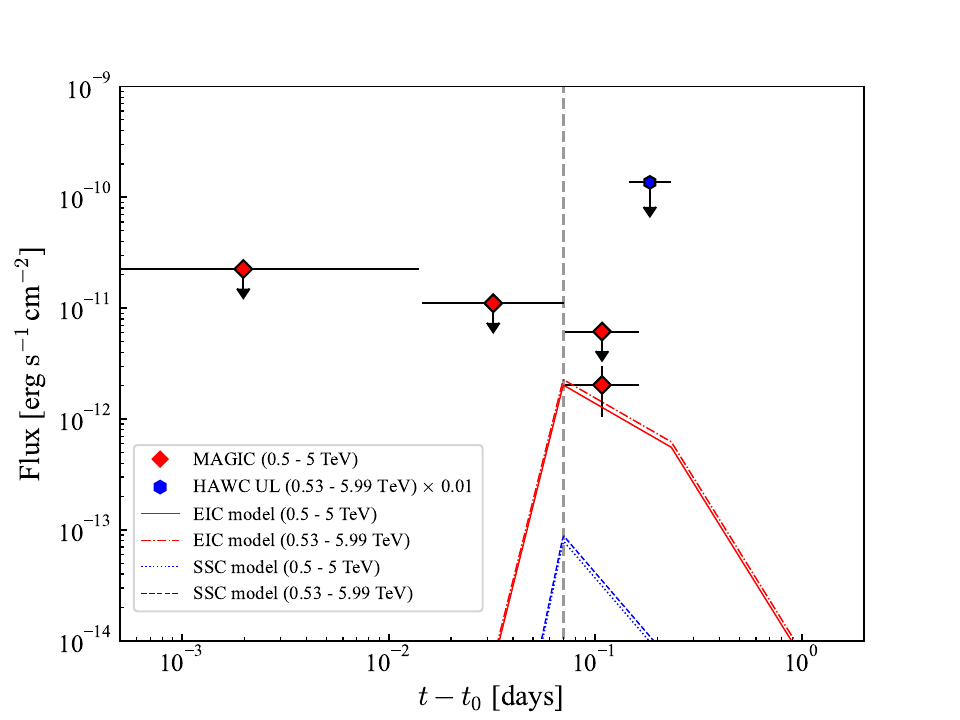}
    \caption{Left: Observations and upper limits of MAGIC and HAWC observatories with the model curves for the EIC and SSC scenarios at energy ranges of 0.5 -- 5 TeV and 0.53 -- 5.99 TeV, respectively.  The best-fit values and representative lower limits found with MCMC simulations using our EIC scenario are reported in Table \ref{Table:ISM_Fit_2}. Right: Similar light curve, but the shock crossing time being constrained to a range from 0.05 -- 0.07 days, as shown in \citet{2019ApJ...883...48L}.}
    \label{fig:GRB160821B}
\end{figure}

\begin{figure}
{\centering
{\includegraphics[width=0.95\linewidth]{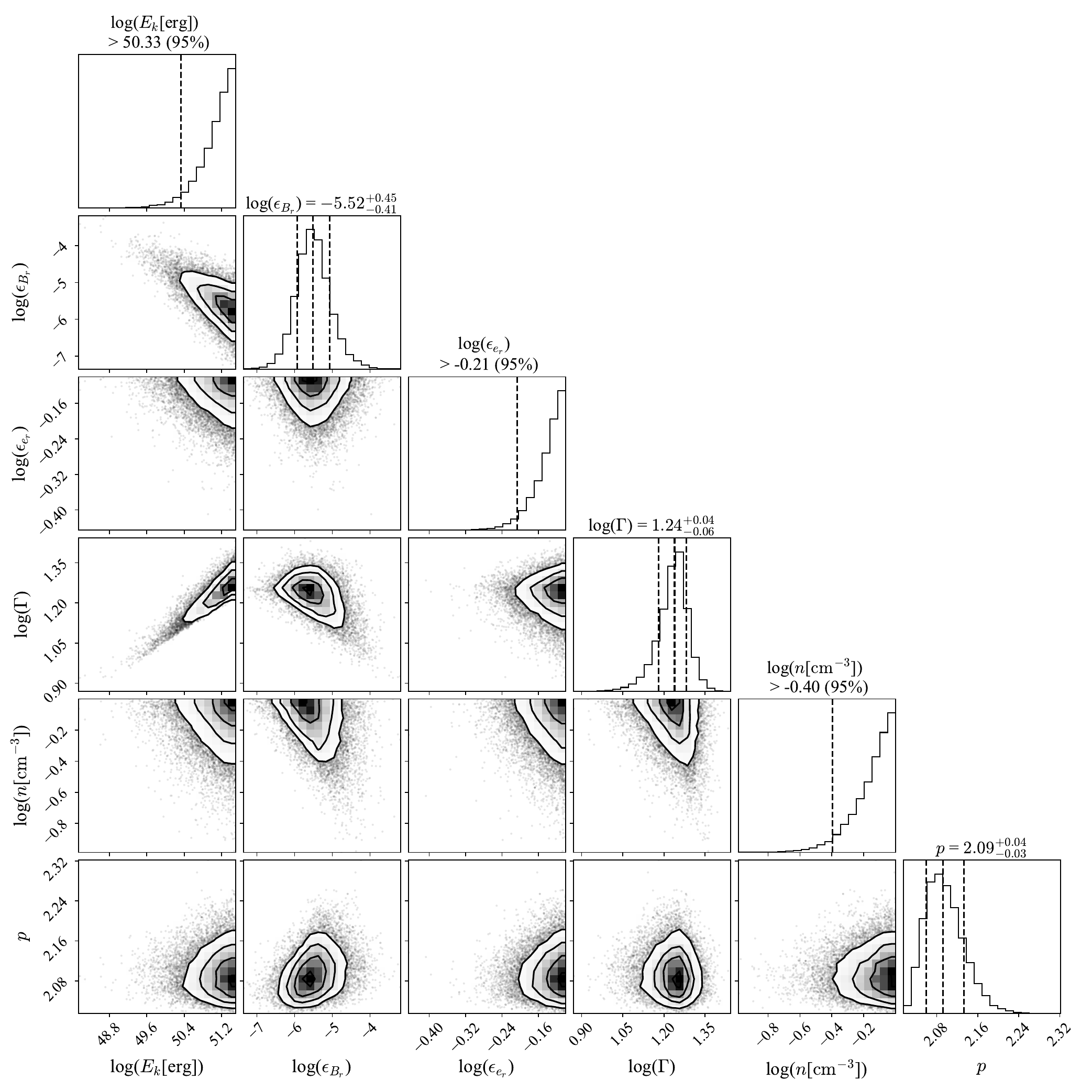}}} 
\caption{Corner plot of the 6-dimensional posterior sampling from the MCMC simulations for GRB 160821B described in the left-hand panel of Figure \ref{fig:GRB160821B}. The diagonal panels show the marginal posterior densities. For well-constrained parameters, the dashed lines denote the best-fit values and central 68\% credible intervals. For parameters affected by prior truncation, only representative 95\% lower credible bounds are shown.
The off-diagonal panels display the joint posterior distributions for all parameter pairs. The contours enclose fixed fractions of posterior probability mass (68\%, 95\%, and 99.7\%), representing Bayesian credible regions of the two-dimensional posteriors. }
\label{mcmc_GRB160821B_1}
\end{figure}

\begin{figure}
{\centering
\resizebox*{0.5\textwidth}{0.35\textheight}
{\includegraphics{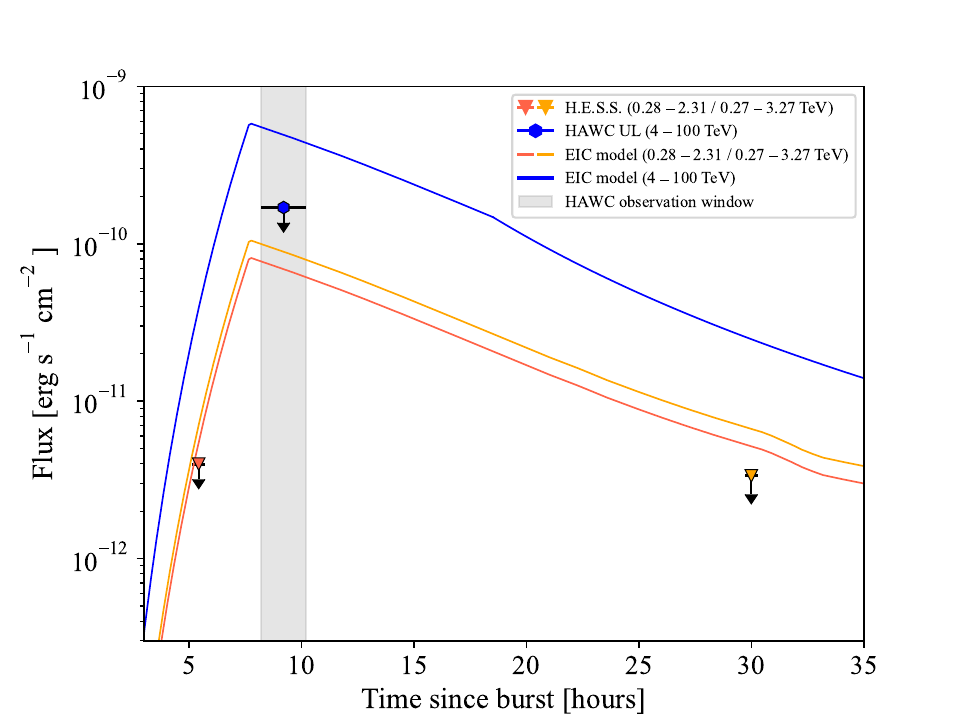}}
\resizebox*{0.5\textwidth}{0.35\textheight}
{\includegraphics{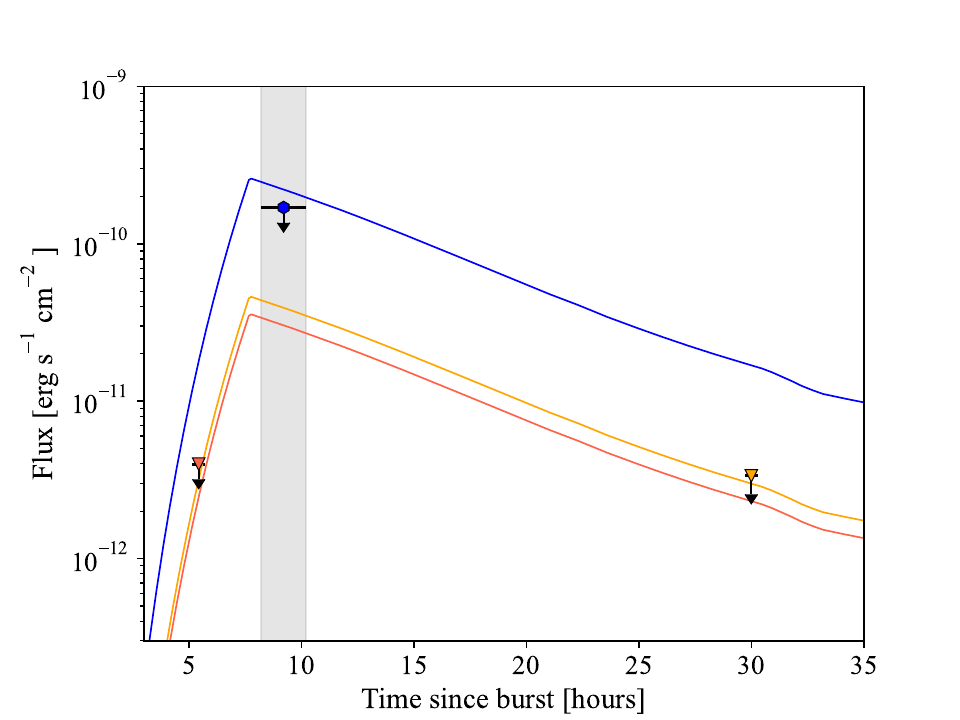}}
\resizebox*{0.5\textwidth}{0.35\textheight}
{\includegraphics{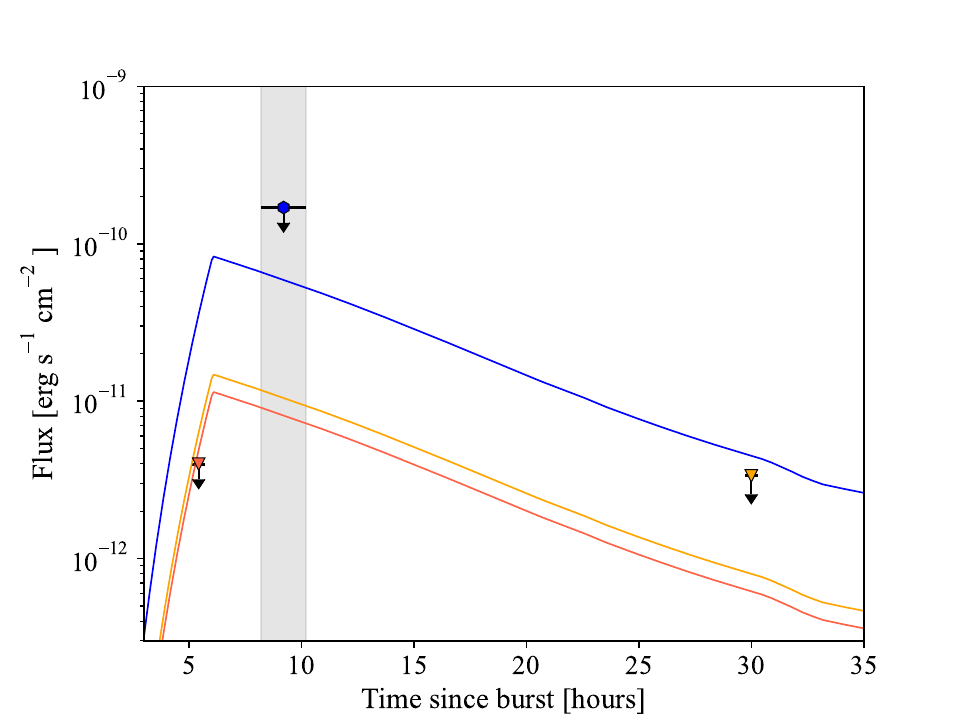}}
\resizebox*{0.5\textwidth}{0.35\textheight}
{\includegraphics{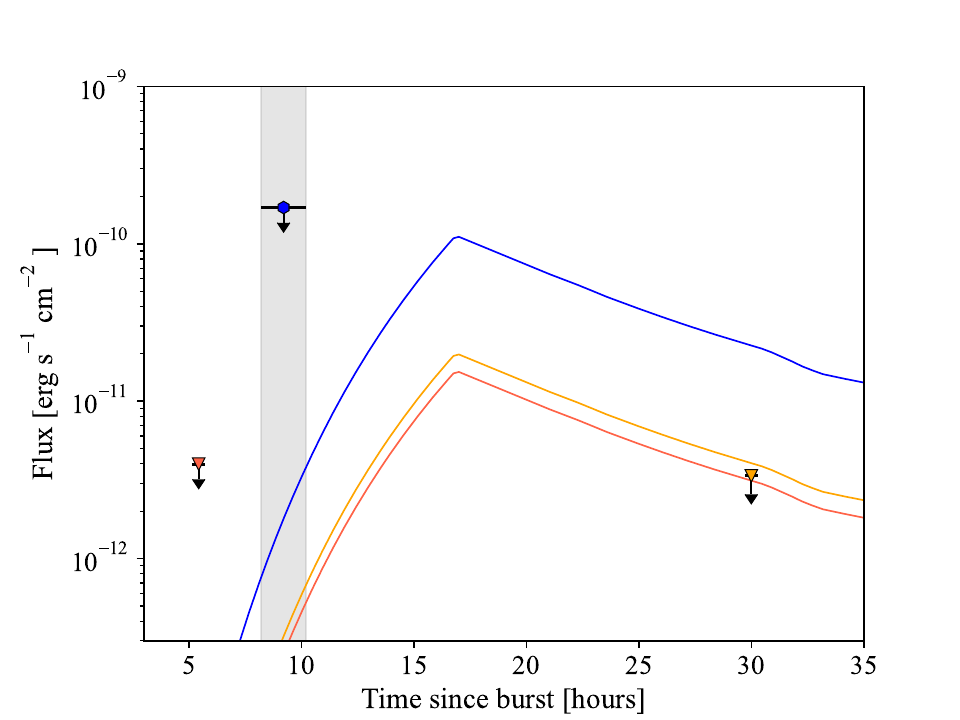}}
} 
\caption{Light curves computed within the EIC framework for four representative parameter sets selected from the excluded region shown in Figure~\ref{param_spac_GRB170817A}.
Left column: The upper panel corresponds to a configuration that exceeds all reported upper limits, whereas the lower panel illustrates a case in which only the earlier HESS upper limit is violated.
Right column: The upper panel shows a solution that only violates the HAWC upper limit, while the lower panel presents a configuration that exceeds only the later HESS upper limit.
The parameter values adopted for each light curve are listed in Table~\ref{Table:ISM_Fit}.}
\label{GRB_170817A}
\end{figure}

\begin{figure}
{\centering
\resizebox*{\textwidth}{0.9\textheight}
{\includegraphics{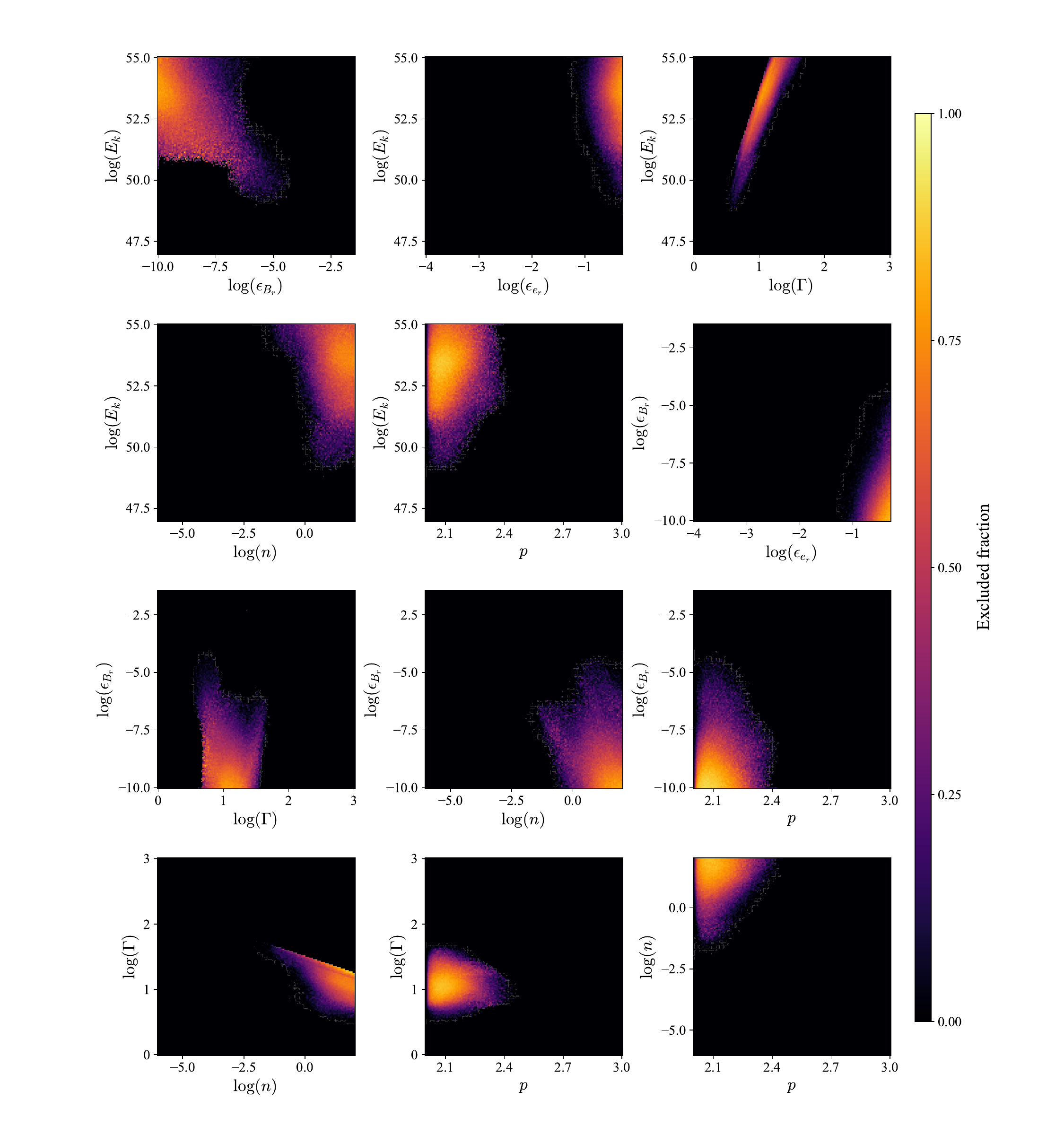}}} 
\caption{2D projections of the explored six-dimensional parameter space ($\rm E_{\rm K},\,\epsilon_{\rm B_r},\,\epsilon_{\rm e_r},\,\Gamma,\,n,\,p$). Each panel show the fraction of parameter combinations yielding fluxes above the HAWC and H.E.S.S. upper limits. The color scale indicates the mean excluded fraction per bin, with brighter regions corresponding to parameter combinations increasingly inconsistent with HAWC and H.E.S.S. upper limits.} 
\label{param_spac_GRB170817A}
\end{figure}

\bsp	
\label{lastpage}

\end{document}